\documentclass[
aip,jcp,
amsmath,amssymb,floatfix,reprint,longbibliography]{revtex4-2}
\usepackage[mathcal]{eucal}
\usepackage{graphicx}
\usepackage{dcolumn}
\usepackage{bm}
\usepackage[utf8]{inputenc}
\usepackage[T1]{fontenc}
\usepackage{mathptmx}
\usepackage{etoolbox} 
\usepackage[caption=false]{subfig}
\usepackage{anyfontsize}
\usepackage{lipsum}
\usepackage{physics}
\usepackage[dvipsnames]{xcolor}
\usepackage{import}
\usepackage{xifthen}
\usepackage{bbm}
\usepackage{accents}
\usepackage[pdftex,plainpages=false,colorlinks=true,citecolor=blue,linkcolor=blue,urlcolor=blue,filecolor=green,bookmarksopen=true]{hyperref}
\usepackage{upgreek}

\DeclareFontFamily{OT1}{mathc}{}
\DeclareFontShape{OT1}{mathc}{m}{it}{<-> mathc10}{}
\DeclareMathAlphabet{\mathcal}{OT1}{mathc}{m}{it}

\newcommand{\dmn}[1]{${#1}\text{-}d$}
\renewcommand{\exp}[1]{\,\mathrm{e}^{#1}} %% EXPONENTIAL
\newcommand{\cc}[1]{#1^{\ast}} %%{\bar{#1}} %% {{#1}^{\ast}} COMPLEX CONJUGATE
\newcommand{\hc}[1]{#1^{\dagger}} %% HERMITIAN CONJUGATE

\newcommand{\pmCustom}[2]{\ooalign{%
		\raisebox{.4\height}{$#1+$}\cr
		\smash{\raisebox{-.3\height}{$#1-$}}\cr}}
\renewcommand{\pm}{\mathbin{\mathpalette \pmCustom \relax}}

\newcommand{\mmax}{\overline{m}} 
\newcommand{\oper}[1]{\hat{#1}}
\renewcommand{\H}[1]{
	\ifthenelse{\isempty{#1}}%
	{\oper{H}}% if #1 is empty
	{\oper{H}_{#1}}% if #1 is not empty
}
\newcommand{\V}[1]{
	\ifthenelse{\isempty{#1}}%
	{\oper{V}}% if #1 is empty
	{\oper{V}_{#1}}% if #1 is not empty
}
\newcommand{\K}[1]{
	\ifthenelse{\isempty{#1}}%
	{\oper{K}}% if #1 is empty
	{\oper{K}_{#1}}% if #1 is not empty
}
\renewcommand{\L}[1]{
	\ifthenelse{\isempty{#1}}%
	{\oper{L}}% if #1 is empty
	{\oper{L}_{#1}}% if #1 is not empty
}
\newcommand{\E}[2]{\oper{E}^{#1}_{#2}}
\renewcommand{\i}{{\mathrm{i}}}
\newcommand{\nn}{\nonumber \\}
\renewcommand*{\dot}[1]{%
  \accentset{\mbox{\large\bfseries .}}{#1}}

\newcommand{\Rmath}{\mathbbm{R}}
\newcommand{\Zmath}{\mathbbm{Z}}

\newcommand{\obs}[1]{\mathcal{#1}}
\newcommand{\obsop}[1]{\hat{\mathcal{#1}\hspace{2pt}}\hspace{-2pt}}

\makeatletter
\def\@email#1#2{%
 \endgroup
 \patchcmd{\titleblock@produce}
  {\frontmatter@RRAPformat}
  {\frontmatter@RRAPformat{\produce@RRAP{*#1\href{mailto:#2}{#2}}}\frontmatter@RRAPformat}
  {}{}
}%
\makeatother

\begin{document}

\title{Effective theory of quantum phases in the dipolar planar rotor chain}

\author{Est\^ev\~ao V.B. de Oliveira}
\affiliation{Department of Physics \& Astronomy, University of Waterloo, Waterloo, Ontario N2L3G1, Canada}
\affiliation{ 
Institute For Quantum Computing, University of Waterloo, Waterloo, Ontario N2L 3G1, Canada
}

\author{Muhammad Shaeer Moeed}
\affiliation{Department of Physics \& Astronomy, University of Waterloo, Waterloo, Ontario N2L3G1, Canada}
\affiliation{ 
Institute For Quantum Computing, University of Waterloo, Waterloo, Ontario N2L 3G1, Canada
}
\affiliation{ Perimeter Institute for Theoretical Physics, Waterloo, Ontario N2L 2Y5, Canada}

\author{Pierre-Nicholas Roy}\email{pnroy@uwaterloo.ca}
\affiliation{ 
Department of Chemistry, University of Waterloo, Waterloo, Ontario N2L 3G1, Canada}
\affiliation{ 
Institute For Quantum Computing, University of Waterloo, Waterloo, Ontario N2L 3G1, Canada
}
\affiliation{ Perimeter Institute for Theoretical Physics, Waterloo, Ontario N2L 2Y5, Canada}

\date{\today}

\preprint{AIP/123-QED}

\begin{abstract}
	In this work, we develop a theoretical description of the collective behavior of interacting dipolar planar rotors by using time independent perturbation theory and a small angle quadratic approximation.
    The ground state properties for both the ordered and disordered quantum phases of the system are directly calculated and analyzed.
    Time-independent perturbation theory is shown to be appropriate for the disordered phase. For the ordered phase, we construct a quadratic approximation based on the stable equilibrium configurations of the dipolar ordering; we show that the inclusion of the quartic terms from the expansion of the potential energy are essential to correct the shift in the energy spectrum due to quantization ambiguities.
    Numerical techniques such as Exact Diagonalization and Density Matrix Renormalization Group are used for the benchmark the quality of both approximations.
\end{abstract}

\maketitle

\section{\label{sec:Introduction}Introduction}

The study of assemblies of interacting confined molecules with rotational degrees of freedom is highly relevant to the fields of chemical, and condensed matter physics, and finds applications in quantum matter and materials science,\cite{endohedral_fullerites_exp, exp_nanotube, aoyagi2014cubic, ma2017quasiphase} including the potential use of arrays of polar molecules as quantum computation devices.\cite{demille2002quantum,Wei_2016}
Experiments have shown that correlated molecular rotations present interesting collective phenomena that would typically be inhibited by direct bulk interactions.
A few representative examples are the confinement of water molecules in crystal structures of beryl,\cite{gorshunov_2016_incipient,kolesnikov2016quantum} cordierite,\cite{belyanchikov2022single} or C$_{60}$ cages in \dmn{1} or \dmn{2} lattices.\cite{exp_nanotube,hou2022synthesis} The crystal structures shield the water molecules from the intermolecular hydrogen bonds, and the now predominant dipole-dipole interactions allow for the appearance of ferroelectricity, tunneling, and ordering behavior at low temperatures.\cite{ma2017quasiphase, aoyagi2014cubic, kolesnikov2016quantum} In this scenario the balance between kinetic energy -- from the molecular rotations -- and the potential energy -- from the dipolar interactions -- determines whether the system will be in a disordered or ordered phase.

\begin{figure}[ht]
    \centering
    \includegraphics[width = 1\columnwidth, trim={0 0 0 0},clip]{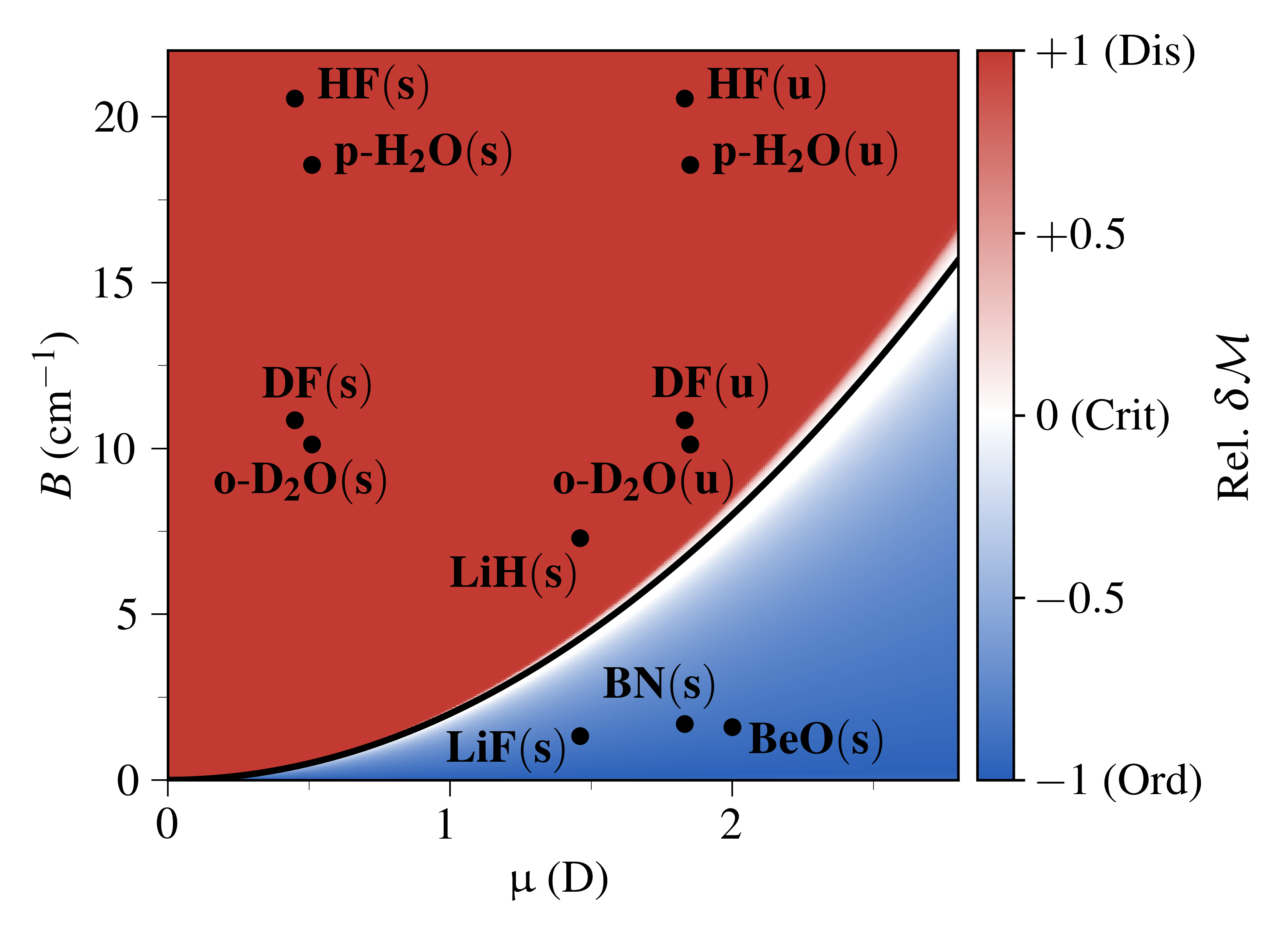}
    \caption[]{\label{fig:B_vs_mu}Phase diagram for endofullerene chains of various guest molecules. The disordered (red), ordered (blue), and critical (white) phases are shown as functions of the rotational constant $B$ and dipole moment $\upmu$ of the confined molecules, considering both shielded (s) and unshielded (u) effective values. 
    The black curve corresponds to the critical pair of $B$ and $\mu^2$. The distance between neighbor dipoles is assumed constant and identical for all the different systems. The critical behaviour is denoted by the relative difference between of the total polarization order parameter calculated analytically and numerically, using the results from Sec. \ref{sec:macro_limit}.
    The values for dipole moments and rotational constants are taken from Refs. \onlinecite{Dolgonos_2014,Serwatka2023endo}.}
\end{figure}

This variety of experimental findings necessitates the development of effective microscopic theories that would provide a deeper understanding of the quantum phases exhibited by confined molecular assemblies.
In this sense, lattices of dipolar rotors have been established as a paradigm for systems of interacting confined molecules with rotational degrees of freedom.\cite{kolesnikov2016quantum,abolins2018quantum, belyanchikov_2020_dielectric,belyanchikov2022single}
As an example, Fig. \ref{fig:B_vs_mu} illustrates the phase diagram for endofullerene chains of various guest molecules with respect to their rotational constant and dipole moment.\cite{Serwatka2023endo} Here, this trade-off between kinetic and potential energies becomes evident.

A larger rotational constant favors the kinetic energy, whereas a larger dipole moment favors the dipole ordering. Hence, this universal behavior, that depends only on the ratio of those two molecular parameters for a given rotor distance, determines whether or not a given molecular assembly exhibits ferro-electric order.
Therefore, a range of confined dipolar molecules can be modeled and analyzed within the same theoretical framework.

Numerically, a plethora of methods have been successfully applied to such systems, each exhibiting inherent advantages and limitations. 
Given the quantum mechanical nature of the model, Exact Diagonalization (ED) is the standard method, although restricted to chains of only a few molecules.\cite{felker2017accurate,felker2017electric, halverson2018quantifying} The exponential scaling of the Hilbert space of the system drastically limits the practical implementation of ED. Larger systems with hundreds of confined molecules can be studied in the ground state using the Density Matrix Renormalization Group (DMRG) method,\cite{white1992density} using Matrix Product State (MPS)\cite{schollwock2011density} to represent the many-body ground state wavefunction. That approach has been applied to study \dmn{1} chains of dipolar linear molecules,\cite{iouchtchenko2018ground} rotating water molecules,\cite{serwatka2022ground,Serwatka2023QPT} water chains in carbon nanotubes,\cite{serwatka2022ferroelectric} various endofullerene peapods,\cite{Serwatka2023endo} and chains of dipolar planar rotors.\cite{Serwatka2024}
Nevertheless, the MPS representation of the ground state wavefunction is limited to one-dimensional chains. 

Another powerful technique to simulate many-body systems in higher dimensional spaces is the Path Integral Monte Carlo (PIMC) method,\cite{ceperley1995path,marx1999path} with the Path Integral Ground State (PIGS) formulation\cite{sarsa2000path,yan2017path} for ground state properties.
This technique has recently been used to calculate ground state angular position and angular momentum properties for linear chains of dipolar planar rotors.\cite{Zhang2024,Moed2024,deOliveira2025}
Such a system is known to present a $(1+1)$-dimensional $\Zmath_2$ symmetry breaking Quantum Phase Transition (QPT) as a fundamental feature.\cite{Serwatka2024,Zhang2024,Moed2024,deOliveira2025}
However, since the configuration space grows exponentially with  system size, an efficient sampling can become unfeasible especially around the critical region.\cite{WOLFF199093, Bonati2018}

In summary, each numerical method applied to model lattices of dipolar rotors has its regime of accuracy and intrinsic limitations. In view of this, the goal of this work is to extend this theoretical study of confined molecular rotations from an analytical perspective. Here, we will formally develop an effective theory to describe both the disordered and ordered regimes of a linear chain of dipolar planar rotors.
The remainder of this paper is organized as follows: in Sec. \ref{sec:Theory} we introduce the system of dipolar planar rotors and apply time independent perturbation theory to describe the disordered quantum phase. We also develop an effective harmonic theory for the ordered quantum phase based on a quadratic expansion.
We report the results in Sec. \ref{sec:results} and discuss our findings. We provide concluding remarks in Sec. \ref{sec:Conclusions} along with a perspective for future work.

\section{Theory}
\label{sec:Theory}

\subsection{The Classical Planar Rotors System}
\label{sec:planar_rotors}
\begin{figure}
  \centering
  \includegraphics[width = 0.9\columnwidth]{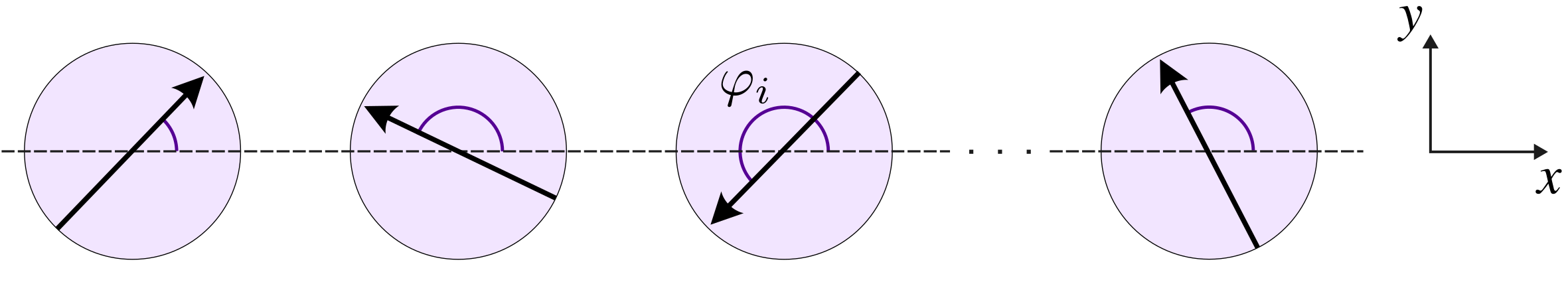}
  \caption{Illustration of the $N$ planar rotor chain in a co-planar arrangement.}
  \label{fig:planar_rotors_chain}
\end{figure}

The system of interest, depicted in Fig. \ref{fig:planar_rotors_chain}, consists of $N$ identical planar rotors with lattice spacing $r$ in a linear chain with co-planar arrangement. Each rotor has angular orientation $\varphi$ in the $xy$-plane and interacts with its nearest neighbors via a dipole-dipole interaction. The state of the $i$-th rotor can be specified by its angular displacement $\varphi_i$ about the $x$-axis, and its angular momentum in the $\vb{z}$ direction $\va{L}_i = I \dot{\varphi}_i \vb{z}$, where $I$ is the moment of inertia.
The rotational kinetic energy of the system is
\begin{align}
	\label{eq:kinetic_energy_rotor_i}
	K = \sum_{i} \frac{I {\dot{\varphi}_i}^2}{2} = \sum_{i} B L_i^2
	,
\end{align}
where $B \equiv \frac{1}{2I}$ is the rotational constant,
and the potential energy due to the dipole-dipole interaction between nearest neighbors is
\begin{align}
	\label{eq:potential_energy_rotors}
	=g \sum_{\langle ij \rangle} \sin{\varphi_i}\sin{\varphi_j} - 2\cos{\varphi_i}\cos{\varphi_j}
	\,,
\end{align}
where $g \propto \frac{\upmu^2}{r^3}$ is the interaction strength. 
Then, the classical Hamiltonian of the system is directly obtained\cite{lemos2018}
\begin{align}
	\label{eq:classical_hamiltonian}
	H = \sum_{i} L_i^2 + g \sum_{\langle ij \rangle} \sin{\varphi_i}\sin{\varphi_j} - 2\cos{\varphi_i}\cos{\varphi_j}
	,
\end{align}
where we will typically adopt units of $\frac{1}{2I} = 1$ (or $\hbar = \frac{1}{2I} = 1$ for the quantum case) throughout the paper.
As mentioned, the system exhibits a $(1+1)$-dimensional $\Zmath_2$ symmetry breaking QPT from an disordered to a ordered phase for some critical interaction strength $g_c \sim 0.5 $.\cite{Serwatka2023endo,Serwatka2024,Zhang2024,Moed2024,deOliveira2025}
In the next sections, we will introduce analytical approaches to describe both quantum phases, motivated by the QPT and phenomenology involved.  

\subsection{Quadratic approximation for the ordered phase}
\label{sec:small_oscillations}

A direct consequence of the invariance of the system Hamiltonian of Eq. \eqref{eq:classical_hamiltonian} with respect to a collective rotation of $\pi$ of all dipoles, the $\Zmath_2$ symmetry, is the appearance of a twofold degenerate ground state in the limit of $g\rightarrow\infty$, with all the rotors in the state $\varphi=0$ or $\varphi=\pi$.\cite{deOliveira2025,MoeedThesis,Serwatka2024}
In fact, the reference states $\va{\varphi} \equiv (\varphi_1, \varphi_2, \dots, \varphi_N) = \va{\theta} \equiv (\theta, \theta, \dots, \theta)$, for $\theta=0,\pi$, are stable equilibrium points of the potential $V(\va{\varphi})$ (see Eq. (S1) from the Supplemental Material).
Then, in the ordered phase regime, it is reasonable to assume that each rotor can be found in a state given by a small displacement $\xi_i \equiv  \varphi_i-\theta$ from its equilibrium state $\theta$. Then, a quadratic approximation of the potential $V(\va{\varphi})$ around the point $\va{\theta}$ gives
\begin{align}
	\label{eq:potential_SOA}
	V
	\approx 
	-2gN + g \sum_{i} \left(2 \xi_{i}^2  + \frac{\xi_{i-1}\xi_i + \xi_{i+1}\xi_i}{2}\right),
\end{align}
In addition, the kinetic energy is simply
\begin{align}
	\label{eq:kinetic_SOA}
	K = \sum_i = \frac{I}{2} \dot{\xi}_i^2 ,
\end{align}
and the total classical Hamiltonian of the system in the quadratic approximation is
\begin{align}
	\label{eq:hamiltonian_SOA}
	H = -2gN +  \sum_{i} \frac{I}{2} \dot{\xi}_i^2 +  g \left(2 \xi_{i}^2  + \frac{\xi_{i-1}\xi_i + \xi_i\xi_{i+1}}{2}\right) .
\end{align}

\subsubsection{Reciprocal lattice and normal modes}
\label{sec:normal_modes}
In order to better analyze the collective behavior of the system, we can write the angular displacement variables $\xi_i$ in terms of the normal modes (NM) (Fourier modes) of the lattice, defining
\begin{align}
	\label{eq:xi_FT}
	\xi_i = \frac{1}{\sqrt{N}} \sum_{j} \phi_j \exp{\i \frac{2\pi}{N}ij}
	\implies
	\phi_j = \frac{1}{\sqrt{N}} \sum_{i} \xi_i \exp{-\i \frac{2\pi}{N}ij}
	.
\end{align}
where $j$ can assume $N$ discrete values corresponding to the first \textit{Brillouin zone} of this one-dimensional crystal.
Also, the displacements are real, implying the condition that
$\phi_{-j} = \cc{\phi}_{j}$
Finally, by writing the kinetic and potential energy in terms of the NM variables (see Eqs. (S3) and (S4)) Eq. \eqref{eq:hamiltonian_SOA} becomes
\begin{align}
	\label{eq:hamiltonian_SOA_NM}
	H =	
	-  2gN	
	+  \frac{I}{2} \sum_j  \dot{\phi}_j \cc{\dot{\phi}}_j +\omega_j^2 \phi_j \cc{\phi}_j
	,
\end{align}
where $\omega_j \equiv \omega_j(g)$ is an explicit function of the interaction strength $g$, and we defined 
$\omega_j(g) = \sqrt{\frac{2g}{I}} \kappa_j $ with $\kappa_j = \sqrt{2 + \cos(\frac{2\pi}{N}j)}$. 

Here, we can already see that in the ordered phase, the system behaves as a set of $N$ decoupled Harmonic Oscillators (HO) with frequency $\omega_j$. For the purpose of rewriting Eq. \eqref{eq:hamiltonian_SOA_NM} in its usual form, we can perform the following change of variables. First, by defining 
\begin{align}
	\label{eq:phi_definition_NM}
	\phi_j &= a_j \exp{\i \omega_j t}+ \cc{a}_{-j} \exp{-\i \omega_j t} ,
\end{align}
with $a_j$ being time independent functions and the relation $\phi_{-j} = \cc{\phi}_{j}$ being preserved, the system Hamiltonian of Eq. \eqref{eq:hamiltonian_SOA_NM} can now be written in the spectral decomposition form (see Eq.(S12)),
\begin{align}
	\label{eq:classical_Hamiltonian_SOA_NM_spec_decomp}
	H =	
	-2gN 
	+ \sum_j \omega_j^2 a_j \cc{a}_j
	.
\end{align}
We can now introduce $Q_j , P_j \in \Rmath$ the generalized coordinates of position and momentum of the system
\begin{align}
	\label{eq:definition_Q_P_generalized}
	Q_j = \sqrt{\frac{1}{2}}\left(a_j \exp{\i \omega_j t} + \cc{a}_{j} \exp{-\i \omega_j t} \right) ,
\end{align}
and $P_j = \dot{Q}_j$, so that the classical Hamiltonian becomes
\begin{align}
	\label{eq:classical_Hamiltonian_SOA_NM_generalized_coordinates}
	H =	- 2 gN 	+ \sum_j \frac{P_j^2}{2} +  \frac{\omega_j^2 Q_j^2}{2} 
	.
\end{align}
Notably, by using Eqs. \eqref{eq:phi_definition_NM} and \eqref{eq:definition_Q_P_generalized} to write $\{Q_j,P_j\}$ in terms of the original coordinates $\{\xi_j,\dot{\xi}_j\}$ (see Eq. (S15)),
it becomes evident that the zeroth mode is directly related to the centroid properties of the system in both canonical variables. Specifically,
\begin{align}
	\label{eq:P0_in_terms_of_Ltotal}
	P_0
	=  \sum_{i} \frac{\dot{\xi}_i}{\sqrt{2N}}  = \sqrt{\frac{2}{N}}\mathcal{L}
	,
\end{align}
relates to the total angular momentum $\mathcal{L} \equiv \sum_i L_i$ of the system.

\subsubsection{Quantization process}
\label{sec:quantization_process}

The quantization of the system will involve the usual second quantization procedure (see Eq. (S20)),
where the commutation between the canonical quantum variables $\{\hat{Q}_j,\hat{P}_j\}$ is substituted by
\begin{align}
	\comm{\hat{Q}_j}{\hat{P}_j} = \i \longrightarrow \comm{\hat{a}_j}{\hc{\hat{a}}_j} = 1,
\end{align}
with $\{\hat{a}_j,\hat{a}_j\}$ being the creation and annihilation operators of the $j$-th mode lattice phonons. Therefore, in the ordered phase the system can be described by a chain of Quantum Harmonic Oscillators (QHO) with Hamiltonian
\begin{align}
	\label{eq:H_OH}
	\H{} = -2gN + \sum_{j} \omega_j \left(\hc{\hat{a}}_j \hat{a}_j + \frac{1}{2}\right).
\end{align}
and, the total ground state energy of the system in the ordered phase (ord) is simply the zero-point energy ($n_j=0$ excitations for all the modes $j$)
\begin{align}
	\label{eq:E0_ordered_phase}
	\obs{E}_{\text{ord}}(g,N)
	\equiv \expval{\H{}}_{\text{ord}}
    = -2gN +  \sqrt{g} \sum_{j=0}^{N-1} \kappa_{j,{\scriptscriptstyle N}}
    ,
\end{align}
where the dependence with $N$ of the terms in the sum is kept explicit.

In addition, the ground state of the system in the ordered phase is a product state of the QHO Gaussian wave functions of all the modes. The ground state position and momentum wave functions for the $j$-th mode are, respectively,
\begin{align}
	\label{eq:qho_wave_func_position}
	\psi_{0}(Q_j)
	= \left(\frac{\omega_j}{\pi }\right)^{\frac{1}{4}} \exp{-\frac{\omega_j Q_j^2}{2}} ,
\end{align}
and
\begin{align}
	\label{eq:qho_wave_func_momentum}
	\psi_{0}(P_j)
	= \left(\frac{1}{\pi \omega_j}\right)^{\frac{1}{4}} \exp{-\frac{P_j^2}{2 \omega_j}} .
\end{align}
As a consequence, most physical quantities of interest for the original system can now be represented in terms of the moments $\expval*{\oper{Q}_j} = \expval*{\oper{P}_j} = 0$, and $\expval*{\oper{Q}_j^2 }=\frac{1}{2\omega_j}$ and $\expval*{\oper{P}_j^2}=\frac{\omega_j}{2}$.
Besides the ground state energy from Eq. \eqref{eq:E0_ordered_phase}, we will focus our attention on the variance of the total angular momentum
\begin{align}
	\label{eq:L2_ordered_phase}
	\obs{L}^2(g,N)
	\equiv
	\expval{\obsop{L}^2}_{\text{ord}}
    = \frac{\expval{P_0^2}}{2} N
    = \frac{\kappa_0}{2}\sqrt{g}N
	,
\end{align}
the total polarization in the $x$-direction,
\begin{align}
	\label{eq:M_ordered_phase}
	\obs{M}_{\text{ord}}(g,N)  &\equiv \expval{\obsop{M}}_{\text{ord}}
    = \cos{\theta}  \left(N - \sum_{j} \expval{\oper{Q}_j^2} \right)
	\nn
	&
	= \cos{\theta}  \left(N - \frac{1}{4\sqrt{g}}\sum_{j=0}^{N-1} \kappa_{j,{\scriptscriptstyle N}}^{-1} \right)
	,
\end{align}
and lastly, the orientational correlation of the chain
\begin{align}
	\label{eq:Corr_ordered_phase}
	\obs{C}_{\text{ord}}(g,N)  &\equiv \expval{\obsop{C}}_{\text{ord}}
    =N + 2\sum_{j} \left(\kappa_{j}^2 - \kappa_0^2\right)\expval{\oper{Q}_j^2}
	\nn
	&
	=N + \frac{1}{2\sqrt{g}}\sum_{j=0}^{N-1} \frac{\kappa_{j,{\scriptscriptstyle N}}^2 - \kappa_0^2}{\kappa_{j,{\scriptscriptstyle N}}}.
\end{align}
A property of interest is the chemical potential, 
which can be directly calculated from Eq. \eqref{eq:E0_ordered_phase} as
\begin{align}
	\label{eq:mu_ord}
	\mu_{\text{ord}}(N) 
	&= \obs{E}_{\text{ord}}(N) - \obs{E}_{\text{ord}}(N-1)
	\nn
	& =-2g + \sqrt{g}\left[\kappa_{{\scriptscriptstyle 1},{\scriptscriptstyle N}} + \sum_{j=0}^{N-2} \left(\kappa_{j,{\scriptscriptstyle N}} - \kappa_{j,{\scriptscriptstyle N-1}}\right)\right].
\end{align}

\subsection{Time independent perturbation theory  for the disordered phase}
\label{sec:perturb_theo_disorder}
In the disordered phase regime, the dipole-dipole interaction can be introduced as a small perturbation of strength $g\ll1$ to the free rotor kinetic energy. For convenience, the quantization procedure for the Hamiltonian of Eq. \eqref{eq:classical_hamiltonian}, will be carried in the angular momentum basis representation $\{\ket{m_i}, \forall m_i \in \Zmath\}$, resulting in
\begin{align}
    \label{eq:hamiltonian_op_quantum_system}
    \H{} = \sum_{i=1}^N \L{i}^2 + g \sum_{\expval{ij}} \V{ij},
\end{align}
with
\begin{align}
	\label{eq:potential_energy_op}
	\V{ij} = - \frac{1}{4} \left( 3\E{+}{i}\E{+}{j}+\E{+}{i}\E{-}{j}+\E{-}{i}\E{+}{j}+3\E{-}{i}\E{-}{j} \right),
\end{align}
the interaction between the $i$-th and $j$-th rotors in terms of raising and lowering operators $\E{\pm}{i}$ in the angular momentum basis.\cite{kleinert1997proper,KOWALSKI2002109,Kastrup_2006,orr2024operator,deOliveira2025} The action of the operators $\L{i}$ and $\E{\pm}{i}$ onto the basis states $\{\ket{m_i}\}$ is\cite{eh_ek_2008,deOliveira2025}
\begin{align} 
  \L{i} \ket{m_i} =  m_i \ket{m_i},
  \qquad
  \E{\pm}{i} \ket{m_i} = \ket{m_i \pm 1} \label{eq:action_L,Ep,Em_ops}.
\end{align}

The unperturbed ground state energy of the system is simply $\obs{E}_0^{(0)}=0$, corresponding to the state $\ket*{\va{0}}$ with $m_i = 0, \, \forall i$. Then, the ground state energy, the variance of total angular momentum, the total polarization, and the orientational correlation in the disordered phase, with corrections up to second order in $g$, become (see Eq. (S32), (S34), (S37) and (S29))
\begin{align}
	\label{eq:E0_disordered_phase}
	\obs{E}_{\text{dis}}(g,N) \equiv \expval{\H{}}_{\text{dis}} = -\frac{5}{8} g^2 N,
	\\
	\label{eq:L2_disordered_phase}
	\obs{L}^2_{\text{dis}}(g,N) \equiv \expval{\obsop{L}^2}_{\text{dis}} = \frac{9}{8} g^2 N
    ,
	\\
	\label{eq:M_disordered_phase}
	\obs{M}_{\text{dis}}(g,N) \equiv \expval{\obsop{M}}_{\text{dis}} = 0
    ,
	\\
	\label{eq:C_disordered_phase}
	\obs{C}_{\text{dis}}(g,N) \equiv \expval{\obsop{C}}_{\text{dis}} = \frac{g}{4} N
    .
\end{align}
Again, we can directly  calculate the chemical potential \textmu \ from Eq. \eqref{eq:E0_disordered_phase} as
\begin{align}
	\label{eq:mu_dis}
	\mu_{\text{dis}}(N) 
	= \obs{E}_{\text{dis}}(N) - \obs{E}_{\text{dis}}(N-1)
	=-\frac{5}{8}g^2.
\end{align}

\section{Results and Discussion}
\label{sec:results}

Results obtained using DMRG are used as a benchmark throughout this section. The site definition introduced in Ref. \onlinecite{Serwatka2024} is adopted here and the ITensor package\cite{fishman2022itensor} is used for the computations. 
The MPS representation of the ground states was optimized using a two-site DMRG algorithm. The Singular Value Decomposition (SVD) threshold was set at $10^{-8}$ and the relative energy cutoff was set at $10^{-8}$.
The local two-body Hilbert space was represented in the angular momentum eigenbasis $\{\ket{m_i}\}$, with $m_i$ integer such that $m_i \in [-\mmax,+\mmax]$, for $\mmax$ some maximum angular momentum. For finite a $N$ number of rotors, the angular momentum distribution follow a binomial distribution\cite{deOliveira2025} with variance $\expval{\obs{L}^2} = \frac{\mmax N }{2}$. Therefore, given the dipole-dipole interaction strength value, the maximum angular momentum can be set according to $\mmax \sim \left\lceil \sqrt{3g} \right\rceil$. For the DMRG simulations $\mmax = 7$ was adopted,  which is sufficient to
obtain converged results within reasonable runtime.

\subsection{\texorpdfstring{$N=2$}{N=2} Planar Rotors Case Study}
\label{sec:N=2_results}

\begin{figure}[ht] 
    \centering
    \includegraphics[width = 1\columnwidth, trim={0 0 0 0},clip]{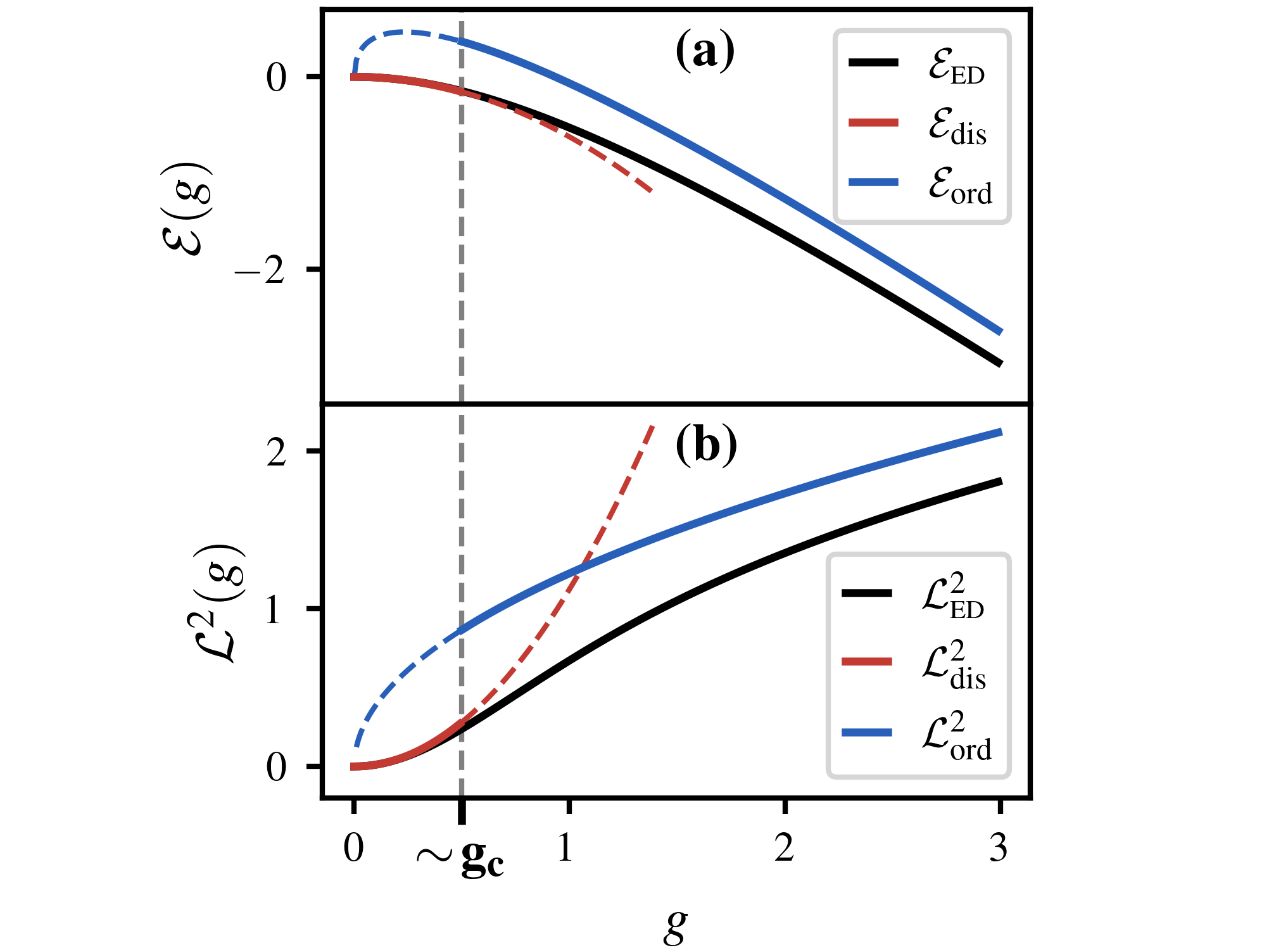}
    \caption[]{\label{fig:N=2_grid}Expectation value of (a) ground state energy and (b) variance of the total angular momentum of the system as a function of the interaction strength $g$ for $N=2$ planar rotors.}
\end{figure}

In order to establish the appropriateness of the approaches 
outlined in Sec. \ref{sec:Introduction}, we use the case of $N=2$ rotors as an illustrative example. The ground state energy and variance of the total angular momentum were calculated for both, the small and large interaction strength regimes, using Eqs. \eqref{eq:E0_ordered_phase} and \eqref{eq:L2_ordered_phase} for $g < g_c$, and, Eqs. \eqref{eq:E0_disordered_phase} and \eqref{eq:L2_disordered_phase} for $g > g_c $. The results were compared to the ED calculations, and are shown in Fig. \ref{fig:N=2_grid}.
At first glance, we see that for both (a) the ground state energy and (b) the variance of the total angular momentum, the asymptotic behavior of the analytical curves closely reproduces the numerical results from ED, up to a constant shift in the regime $g>g_c$.
\begin{figure}[htb] 
    \centering
    \includegraphics[width = 1\columnwidth, trim={0 0 0 0},clip]{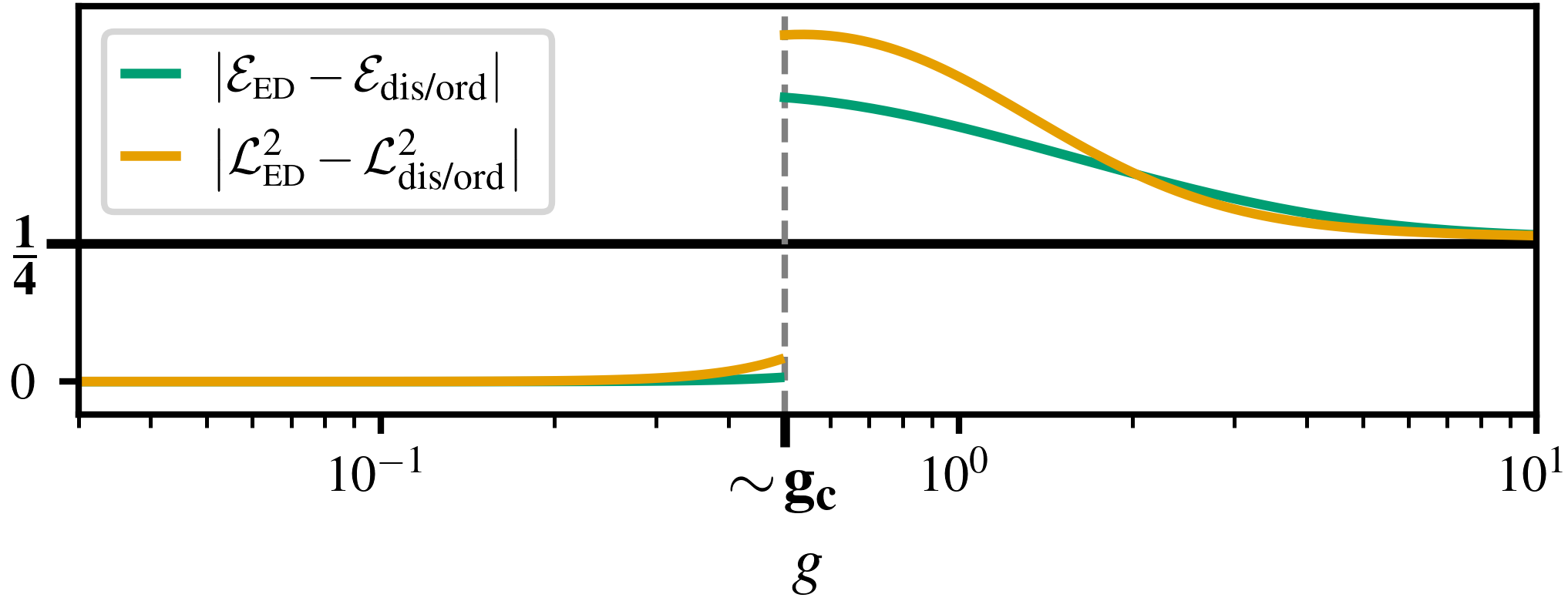}
    \caption[]{\label{fig:N=2_rel_diff}Absolute value of the difference between system properties calculated numerically, using Exact Diagonalization (ED), and analytically.}
\end{figure}

The absolute value of the difference between the analytical and numerical results are shown in Fig. \ref{fig:N=2_rel_diff}, from which it becomes evident that the aforementioned shift converges to $\frac{1}{4}$, as $g$ increases.
This shift in the spectrum arises from the ambiguities present in the quantization process, when mapping the quantum equivalence principle in flat space into laws in curved or constrained spaces.\cite{kleinert1990path,kleinert1993quantum,kleinert1997proper,dolan2021constrained}
The quadratic expansion procedure, that takes Eq. \eqref{eq:potential_energy_rotors} to Eq. \eqref{eq:potential_SOA}, imposes a linearization of the potential energy of Eq. \eqref{eq:potential_energy_rotors} in terms of the canonical variables. This necessarily implies a quantization in $\Rmath\!^1$, not affected by any constraint on the configuration space, unlike the quantization in the compact $S^1$, that results in the operator of Eq. \eqref{eq:potential_energy_op}.

\subsubsection{Exact solution with Mathieu functions}
\label{sec:Mathieu_eqs}

To proceed with the examination of the shift in the spectrum due to the differences between the quantization in flat and curved spaces, it is instructive to consider the system Hamiltonian in the NM variables while preserving the non-linear form of the potential energy from Eq. \eqref{eq:potential_energy_rotors}. For a system with $N=2$ planar rotors, the lattice transformation introduced in sec. \ref{sec:normal_modes}, gives $\phi_j = \frac{\varphi_1 + e^{\i \pi j} \varphi_2}{\sqrt{2}}$, for $j=0,1$. Its direct application to the quantum Hamiltonian of the system in its exact form converts the Schr\"odinger Equation into a set of uncoupled Mathieu Equations of the NM. This leads to
\begin{align}
    \label{eq:Mathieu_Eq}
	\pdv[2]{\psi_j}{\Phi_j} + \left[ 2\obs{E}_j -\frac{\omega_j^2}{4} \cos(2\Phi_j) \right] \psi_j = 0
	,
\end{align}
for $\Phi_j \equiv \frac{\sqrt{2}\phi_j + \pi}{2}$.
In this scenario the NMs are already decoupled by construction, i.e. no quadratic approximation imposed. 

It has been shown that an asymptotic expansion of the Mathieu functions leads to a ground state energy per mode of the form\cite{D_Frenkel_2001, wilkinson_approximate_2018}
\begin{align}
	\obs{E}_j 
	&= -\frac{\omega_j^2}{8} + \frac{\omega_j}{\sqrt{2}} -\frac{1}{8}
	.
\end{align}
This results in a total energy equal to $\obs{E}_{\text{ord}}$ from Eq. \eqref{eq:E0_ordered_phase} plus $\frac{1}{8}$ per mode, in accordance with what is seen in Fig. \ref{fig:N=2_grid}. Therefore, it becomes evident (at least for the system of $N=2$ rotors) that the problem due to quantization disparities between flat and curved spaces must be addressed in the quadratic approximation for the regime $g>g_c$. More precisely, the $\frac{1}{4}$ total shift correction should be taken into account in Eqs. \eqref{eq:E0_ordered_phase} and \eqref{eq:L2_ordered_phase}. In the next section we will examine that problem for the $N>2$ case.

\subsection{\texorpdfstring{$N>2$}{N>2} Planar Rotors Scenario}

In order to address the case of many rotors it is interesting to first focus our attention on the wave function in the momentum space for the ordered phase, Eq. \eqref{eq:qho_wave_func_momentum}. For the ground state of the zero-th mode ($n_0=0$), and using Eq. \eqref{eq:P0_in_terms_of_Ltotal}, we have
\begin{align}
	\label{eq:qho_wave_func_momentum_ground_0th_mode}
	\psi_{0}(\obs{L})
	= \left(\frac{1}{2\pi \sqrt{3g}}\right)^{\frac{1}{4}} \exp{-\frac{\obs{L}^2/N}{2\sqrt{3g}}}.
\end{align}
This indicates that the distribution of total angular momentum follows a normal distribution with mean $\expval{\obs{L}}=0$ in the ordered phase, as already shown in Ref. \onlinecite{deOliveira2025}. In addition, we know from Sec. \ref{sec:N=2_results} that the derived properties $\obs{E}_{\text{ord}}$ and $\obs{L}^2_{\text{ord}}$ have correct asymptotic behavior, up to a constant. 
Nevertheless, although exhibiting the same logic as the previous case, the system with $N>2$ rotors should be treated with extra circumspection for a couple of fundamental reasons. First, the exact determination of the $\frac{1}{8}$ shift per degree of freedom through Mathieu functions is unique to the $N=2$. Second, the proof of concept of the theory developed here relies in the comparison between the physical quantities calculated analytically and numerically. While the former implicitly assumes a system in the macroscopic limit, the latter presents computational limitations to reach that regime. In the next steps we will scrutinize both issues.

\subsubsection{Higher order interactions}
\label{sec:quartic_interactions}

In Section \ref{sec:perturb_theo_disorder}, time independent perturbation theory was used to approximate observables up to second order in $g$. Since $g<g_c \ll 1$ in the disordered phase, higher order terms are not essential to capture the physics of that regime.
Nonetheless, the same should not be assumed in the ordered regime, where the expansion is performed on the variables of the potential energy based on some reference state. Therefore, the contribution of higher order terms and their asymptotic behavior with $g$ should be carefully analyzed. Introducing the higher order terms from the Taylor expansion of the potential energy the Hamiltonian of the system becomes
\begin{align}
    H = H_{\text{HO}} + g\sum_{\alpha >2} \{\phi^{\alpha}\},
\end{align} 
where $H_{\text{HO}}$ is given by Eq. \eqref{eq:H_OH} and $\{\phi^{\alpha}\}$ stands for the contribution of all $n$-th order terms in the expansion, for $n$ even. This contribution is of type 
\begin{align}
	\label{eq:phi_n_contribution}
	\{\phi^{\alpha}\} \propto \sum_{j_1} \dots \sum_{j_{\alpha}} \phi_{j_1} \phi_{j_2} \dots \phi_{j_{\alpha}},
\end{align}
where each $j_i$ stands for a distinct NM. As shown for the $N=2$ case, $j_i = j$ for all $i$, and $\{\phi^{\alpha}\}$ simply reduces to the $\alpha$-th order terms of the expansion of the cosine function from the Mathieu Equations in \eqref{eq:Mathieu_Eq}.

The same is not true for the case where $N>2$. Although the exact calculation of the contribution of $\{\phi^{\alpha}\}$ can still be tractable by means of a diagrammatic perturbation theory approach,\cite{altland2010condensed} for the scope of this paper a simple analysis of the asymptotic behavior of the $\alpha$-th order terms suffices to describe their effects. Introducing $\{\phi^{\alpha}\}$ as a perturbation in the ordered phase, the fluctuations due to this quantity can be replaced by its average contribution. As a result, it can be shown (see Eqs. (S30) and (S31)) that a first order expansion would give corrections in the ground state energy of
\begin{align}
	\label{eq:1st_order_corrections_phi_n_energy}
	\obs{E}^{(1)}_j \propto g^{\frac{4-\alpha}{4}},
\end{align}
making it clear that all but the $\alpha=4$ higher order terms vanish as $g$ increases. In addition, the corrections up to first order in the ground state ket would be of the form
\begin{align}
	\label{eq:first_order_corrections_phi_n_psi}
	\ket*{\psi_j^{(1)}} \propto g^{\frac{2-\alpha}{4}} \sum_{n_j \neq 0} \lambda_{n_j} \ket{n_j},
\end{align}
for $\lambda_{n_j}$ some probability amplitude, with no $g$ dependence. 

This implies that any physical observable with general form $\oper{O} = g^{\alpha} \, f(\oper{a},\hc{\oper{a}})$ will have expectation value in which the contribution of higher order terms is, at best, proportional to $g^{\alpha+\frac{2-\alpha}{4}}$. For instance, for both  $\obs{E}_{\text{ord}}$ and $\obs{L}^2_{\text{ord}}$, from Eqs. \eqref{eq:E0_ordered_phase} and \eqref{eq:L2_ordered_phase} respectively, $\alpha = \frac{1}{2}$ indicating that $\alpha=4$ is the only non-vanishing higher order term, with constant contribution of precisely $\frac{1}{8}$ per $N$ degree of freedom. In addition, for $\obs{C}_{\text{ord}}$ and $\obs{M}_{\text{ord}}$, from Eqs. \eqref{eq:Corr_ordered_phase} and \eqref{eq:M_ordered_phase}, $\alpha = -\frac{1}{2}$, and all the terms of order $\alpha>2$ vanish with $g$. This shows that this mean-field--like approach suffices to capture the physics of the system in ordered phase equilibrium, and that the interactions due to quartic terms gives the correction expected, as seen in the results shown in Figs. \ref{fig:N=150_grid} and \ref{fig:N=150_rel_diff} further. This description of ordered phase of the system in terms of a Gaussian formulation differs from the usual mean-field theory approaches for dipolar rotors, that depend on the direct numerical integration over the degrees of freedom of the exact system Hamiltonian.\cite{abolins2018quantum, BoLiu}

\subsubsection{The macroscopic limit}
\label{sec:macro_limit}

Now, a few points should be made in order to address the nuances of the comparison between physical quantities calculated analytically and numerically. 
While our analytical results correspond to the thermodynamic limit, computational results naturally contain finite size effects. Therefore, we need to ensure computational convergence by using a sufficiently large system for each of the observables under consideration. It is also instructive to observe how physical quantities of interest change with system size in this context.

\begin{figure}[ht]
    \centering 
    \includegraphics[width = 1\columnwidth, trim={0 0 0 0},clip]{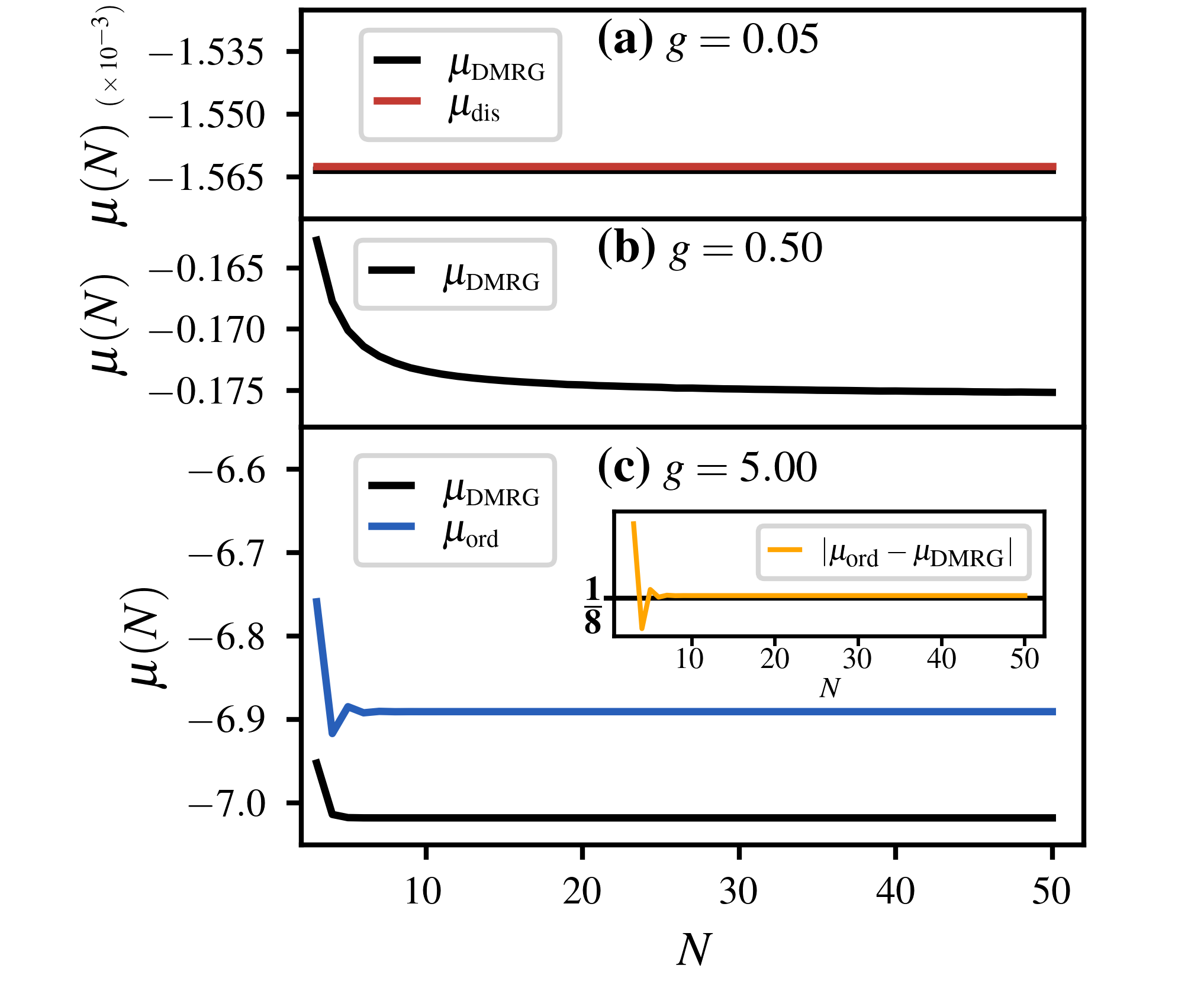}
    \caption[]{\label{fig:mu_vs_N}Chemical potential as a function of the system size for the regimes (a) $g \ll g_c$, (b) $g \sim g_c$, and (c) $g \gg g_c$, calculated numerically using the DMRG algorithm and analytically, using Eqs. \eqref{eq:mu_dis} and \eqref{eq:mu_ord} for the insets (a) and (c) respectively.}
\end{figure}

Thus, the chemical potential of the system in the regimes $g \ll g_c$, $g \sim g_c$, and $g \gg g_c$ was calculated numerically using the DMRG algorithm for different system sizes. The results are shown in Fig. \ref{fig:mu_vs_N}, along with the analytical chemical potential for the ordered and disordered phases, from Eqs. \eqref{eq:mu_ord} and \eqref{eq:mu_dis} respectively.
Here, we can immediately observe the critical behavior of the system for $g=0.5 \sim g_c$, where the chemical potential takes longer to reach a plateau due to long range correlations.\cite{iouchtchenko2018ground,serwatka2022ground,Serwatka2023QPT,Serwatka2024}
For this regime, the theory derived in Sec. \ref{sec:Introduction} presuming phase equilibrium does not hold. For the $g=0.05\ll g_c$ regime, the chemical potential from Eq. \eqref{eq:mu_dis} is in agreement with the DMRG results, and it is constant for all system sizes. 
This indicates that the thermodynamic limit is rapidly reached, as anticipated given the predominance of the free rotational kinetic energy and the resulting linear dependence on $N$ of $\obs{E}_{\text{dis}}$ from Eq. \eqref{eq:E0_disordered_phase}. 

The same can also be observed for the opposite regime, $g=5.00 \gg g_c$, where the constant chemical potential is achieved with only a few rotors. Here, the chemical potential is also affected by the constant shift, and by taking the absolute difference between the chemical potential from analytical and numerical results (inset of Fig. \ref{fig:mu_vs_N} (c)) we can determine the $\frac{1}{8}$ contribution per $N$ degree of freedom. Taking that into account, we can now legitimate the assumption that in this scenario, the system approaches the thermodynamic limit, so we can eliminate the dependence with $N$, by taking the summations in Eqs. \eqref{eq:E0_ordered_phase}, \eqref{eq:L2_ordered_phase}, \eqref{eq:M_ordered_phase}, and \eqref{eq:Corr_ordered_phase} to the continuum limit, making the substitution
\begin{align}
	\frac{1}{N}\sum_{j=0}^{N-1} \longrightarrow \frac{1}{2\pi}\int_{0}^{2\pi}.
\end{align}

\begin{figure}[ht]
    \centering 
    \includegraphics[width = 1\columnwidth, trim={0 0 0 0},clip]{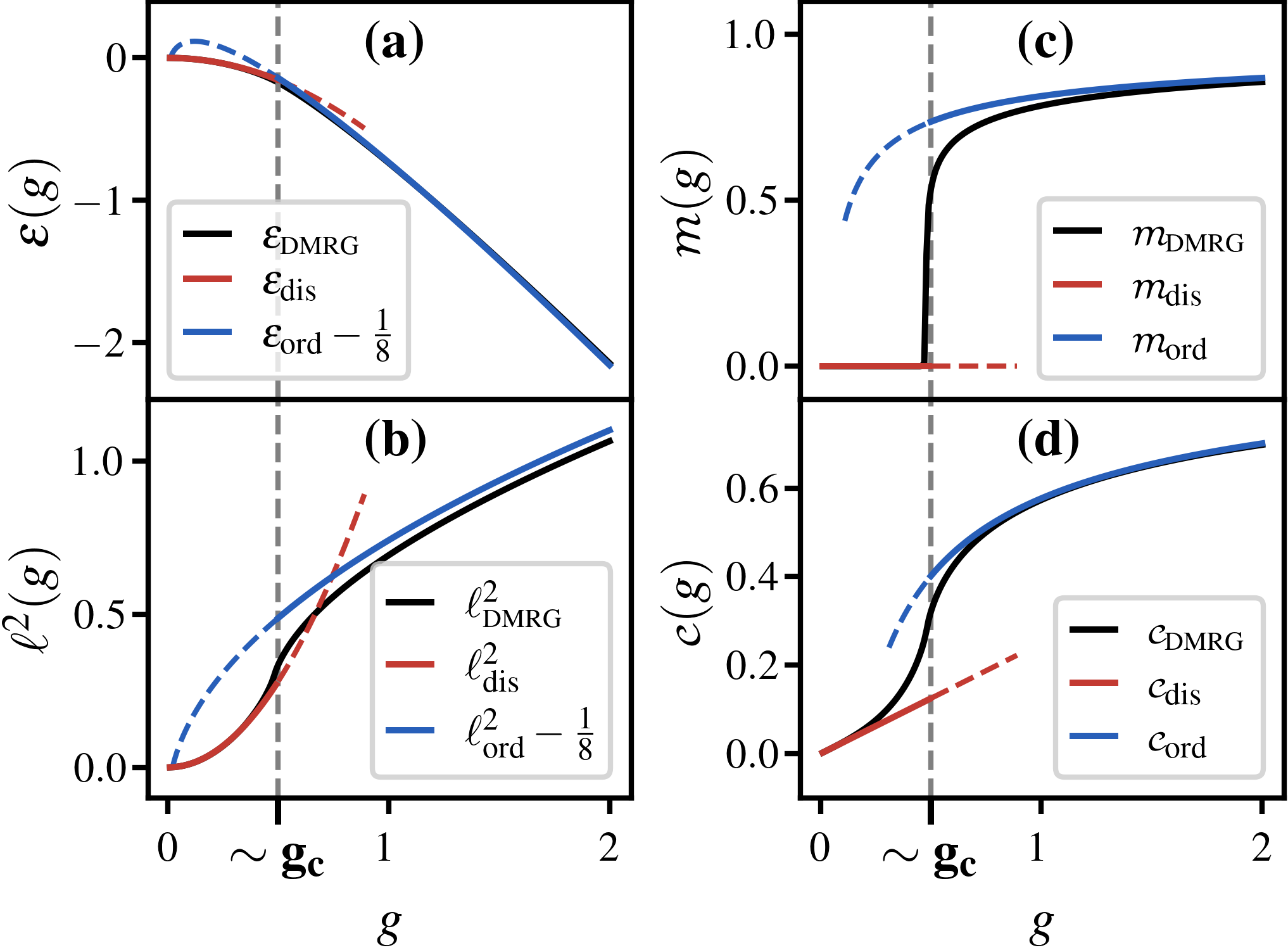}
    \caption[]{\label{fig:N=150_grid}Expectation value of (a) ground state energy, (b) variance of the total angular momentum, (c) polarization and (d) orientational correlation for $N=150$ planar rotors as a function of the interaction strength $g$.}
\end{figure}

\begin{figure}[ht]
    \centering
    \includegraphics[width = 1\columnwidth, trim={0 0 0 0},clip]{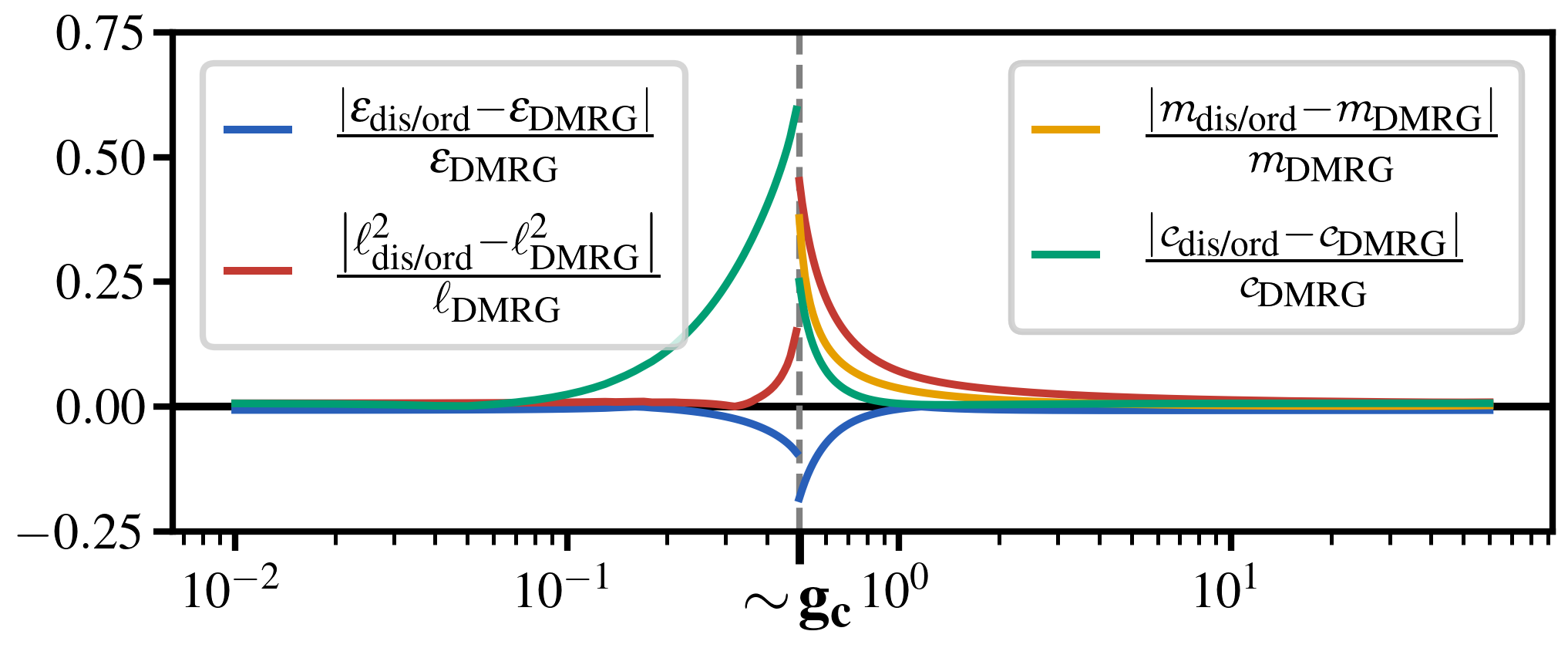}
    \caption[]{\label{fig:N=150_rel_diff}Relative difference between system properties calculated analytically abd the numerical DMRG results.}
\end{figure}

Then, the physical quantities of interest become
\begin{align}
	\label{eq:E0_ordered_phase_N_big}
	\varepsilon_{\text{ord}}
	&= -2g +  \frac{2\obs{I}_{e}\!\left(\textstyle \frac{2}{3}\right)}{\pi} \sqrt{3g}
    ,
	\\
	\label{eq:L2_ordered_phaseN_big}
	\ell^2_{\text{ord}}
	&= \frac{\sqrt{3g}}{2}
	,
	\\
	\label{eq:M_ordered_phaseN_big}
	\obs{m}_{\text{ord}} 
    &= \cos{\theta}  \left(1 - \frac{ \obs{I}_{k}\!\left(\textstyle \frac{2}{3}\right)}{2\pi \sqrt{3g}} \right)
	,
	\\
	\label{eq:Corr_ordered_phaseN_big}
	\obs{c}_{\text{ord}} 
	&= 1- \frac{\sqrt{3}\left(\obs{I}_{e}\!\left(\textstyle \frac{2}{3}\right)-\obs{I}_{k}\!\left(\textstyle \frac{2}{3}\right)\right)}{\pi\sqrt{g}},
\end{align}
where $\obs{I}_{k}$ and $\obs{I}_{e}$ are the complete elliptic integrals of first and second kind, respectively.\cite{abramowitz1965handbook}
The asymptotic behavior of those quantities with respect to $g$ is compared to the results calculated numerically using DMRG, as shown in Fig. \ref{fig:N=150_grid}, as well as their relative difference with respect to the same numerical findings, presented in Fig. \ref{fig:N=150_rel_diff}.
For Figs. \ref{fig:N=150_grid} (a) and (b), the contribution due to $\{\phi^4\}$ terms is taken into account, whereas for the $\obs{m}_{\text{ord}}$ and $\obs{c}_{\text{ord}}$ quantities from Figs. \ref{fig:N=150_grid} (c) and (d), this contributions is absent, as anticipated in Sec. \ref{sec:quartic_interactions}. The plots from Fig. \ref{fig:N=150_rel_diff} show that, aside from the critical region, the analytical predictions are in agreement with the numerical benchmarks. 
Here, it is important to emphasize that the system size, both in terms of the number of rotors and dimension of the basis state representation, does not impose a limitation on the direct calculation of physical properties. The same cannot be said about the numerical methods mentioned in Sec. \ref{sec:Introduction}, in which either the viability or the convergence are contingent upon the truncation of the Hilbert space of the system at the outset.

\section{Concluding remarks}
\label{sec:Conclusions}

We introduced analytical approaches to study the ground state properties of dipolar planar rotors in a linear chain. In the disordered regime ($g<g_c$) we used time-independent perturbation theory, where the dipole-dipole interaction was treated as a perturbation to the free rotor Hamiltonian. 
Corrections to the ground state energy of the system were calculated up to second order in $g$. Also, corrections up to first order in $g$ to the ground state wavefunction were calculated, enabling the calculation of expectation values such as the variance of the total angular momentum, total polarization in the $x$-direction, and orientational correlation properties. The agreement between the calculated ground state properties and the ED results implies that treating the dipole-dipole interaction as a small perturbation is sufficient to describe the system deep in the disordered phase. 

In the ordered regime ($g>g_c$) we applied a quadratic approximation to the dipole-dipole potential energy using the stable equilibrium geometry as the reference point for the expansion. The subsequent lattice transformation of the system Hamiltonian allowed for a description of the system in terms of decoupled QHOs of the normal modes. 
The comparison between energetic properties calculated analytically and numerically reveals a constant shift of $\frac{1}{8}$ per degree of freedom in the spectrum due to the differences of the quantization process in flat and curved spaces. We showed that introducing the quartic terms of the expansion of the potential energy as a perturbation accounts for this correction. Furthermore, the variance of the total angular momentum, the total polarization in the $x$-direction, and the orientational correlation can be directly calculated for the ordered phase with excellent agreement with the DMRG numerical results.

In the spirit of Fig. \ref{fig:B_vs_mu}, we can infer that the analytical framework developed here is applicable for a wide range of dipolar planar rotors system for both the disordered and ordered phases, provided that the $B$ constant, rotor spacing, and effective dipole moment of the trapped species yield an interaction strength, $g$,  away from the critical value, $g_c$.
We stress that the regime of validity of the method relies on the perturbative approximations around reference states, constituting an effective theory for the disordered and ordered phases of these correlated rotor systems. Although unable to describe any behavior in the critical region by construction, this formalism represents a very useful analytical tool to interpret experimental realizations that place the system deep into one of the two phases.

Also, for the ordered regime, it is important to notice that the application of the mean-field-theory locally, with the plain linearization of the Hamiltonian made no distinction regarding the geometry or constraints of the quantized space. The Gaussian integration of the $\{\phi^4\}$ terms suggests that applying the theory globally could intrinsically addresses these issues, and therefore, the development of a proper $\phi^4\!$-theory for the system will be the subject of future work. In addition, the treatment of other quantities, such as the entanglement entropy and response properties, are within the scope of the present analytical framework and will be considered in future studies.

\section*{Supplemental Material}
See the supplemental material for additional results and derivations that were omitted from the main text for the sake of brevity.

\begin{acknowledgments}
This research was supported by the Natural Sciences and Engineering Research Council (NSERC) of Canada (RGPIN-03725-2022), the Ontario Ministry of Research and Innovation (MRI), the Canada Research Chair program (950-231024), the Digital Research Alliance of Canada, and the Canada Foundation for Innovation (CFI) (project No. 35232).
\end{acknowledgments}

\section*{Data Availability Statement}

The data that support the findings of this study are available
from the corresponding author upon reasonable request.

\bibliography{references}

@book{lemos2018,
  title     = {Analytical Mechanics},
  author    = {Lemos, N.A.},
  isbn      = {9781108416580},
  lccn      = {2018011108},
  url       = {https://books.google.ca/books?id=JbViDwAAQBAJ},
  year      = {2018},
  publisher = {Cambridge University Press},
  doi       = {10.1017/9781108241489}
}

@article{abolins2018quantum,
  title={Quantum phases of dipolar rotors on two-dimensional lattices},
  author={Abolins, Brendan P and Zillich, Robert E and Whaley, K Birgitta},
  journal={J. Chem. Phys.},
  volume={148},
  number={10},
  pages={102338},
  year={2018},
  publisher={AIP Publishing},
  doi={10.1063/1.5005522}
}

@article{sarsa2000path,
  title={A path integral ground state method},
  author={Sarsa, A and Schmidt, KE and Magro, WR},
  journal={J. Chem. Phys.},
  volume={113},
  number={4},
  pages={1366--1371},
  year={2000},
  publisher={American Institute of Physics},
  doi={10.1063/1.481926}
}

@article{yan2017path,
  title={Path integral Monte Carlo ground state approach: Formalism, implementation, and applications},
  author={Yan, Yangqian and Blume, D},
  journal={J. Phys. B: At. Mol. Opt. Phys.},
  volume={50},
  number={22},
  pages={223001},
  year={2017},
  publisher={IOP Publishing},
  doi={10.1088/1361-6455/aa8d7f}
}

@article{marx1999path,
  title={Path integral simulations of rotors: theory and applications},
  author={Marx, Dominik and M{\"u}ser, Martin H},
  journal={J. Phys.: Condens. Matter},
  volume={11},
  number={11},
  pages={R117--R155},
  year={1999},
  publisher={Institute of Physics},
  doi={10.1088/0953-8984/11/11/003}
}

@article{ceperley1995path,
  title = {Path integrals in the theory of condensed helium},
  author = {Ceperley, D. M.},
  journal = {Rev. Mod. Phys.},
  volume = {67},
  issue = {2},
  pages = {279--355},
  numpages = {0},
  year = {1995},
  month = {Apr},
  publisher = {American Physical Society},
  doi = {10.1103/RevModPhys.67.279},
  url = {https://link.aps.org/doi/10.1103/RevModPhys.67.279}
}

@article{orr2024operator,
  title={Operator Formulation of Feynman Path Centroid Dynamics for Rotations},
  author={Orr, Lindsay and Roy, Pierre-Nicholas},
  journal={J. Phys. Chem. A},
  volume={128},
  number={17},
  pages={3419--3433},
  year={2024},
  publisher={ACS Publications},
  doi={10.1021/acs.jpca.3c08201}
}

@article{KOWALSKI2002109,
title = {Exotic behaviour of a quantum particle on a circle},
journal = {Phys. Lett. A},
volume = {293},
number = {3},
pages = {109-115},
year = {2002},
issn = {0375-9601},
doi = {https://doi.org/10.1016/S0375-9601(02)00007-5},
url = {https://www.sciencedirect.com/science/article/pii/S0375960102000075},
author = {K Kowalski and J Rembieliński}
}

@article{deOliveira2025,
    author = {V. B. de Oliveira, Estevao and Moeed, Muhammad Shaeer and Roy, Pierre-Nicholas},
    title = {Path integral Monte Carlo in the angular momentum basis for a chain of planar rotors},
    journal = {J. Chem. Phys.},
    volume = {163},
    number = {14},
    pages = {144112},
    year = {2025},
    month = {10},
    issn = {0021-9606},
    doi = {10.1063/5.0284612},
    url = {https://doi.org/10.1063/5.0284612}
}

@Article{kolesnikov2016quantum,
  author    = {Kolesnikov, Alexander I and Reiter, George F and Choudhury, Narayani and Prisk, Timothy R and Mamontov, Eugene and Podlesnyak, Andrey and Ehlers, George and Seel, Andrew G and Wesolowski, David J and Anovitz, Lawrence M},
  title     = {Quantum tunneling of water in beryl: a new state of the water molecule},
  journal   = {Phys. Rev. Lett.},
  year      = {2016},
  volume    = {116},
  number    = {16},
  pages     = {167802},
  publisher = {APS},
  doi       = {10.1103/physrevlett.116.167802}
}

@Article{felker2017accurate,
  author    = {Felker, Peter M and Ba{\v{c}}i{\'c}, Zlatko},
  title     = {Accurate quantum calculations of translation-rotation eigenstates in electric-dipole-coupled H2O@ C60 assemblies},
  journal   = {Chem. Phys. Lett.},
  year      = {2017},
  volume    = {683},
  pages     = {172--178},
  publisher = {Elsevier},  doi       = {10.1063/1.4953180}
}

@article{felker2017electric,
  title={Electric-dipole-coupled H2O@ C60 dimer: Translation-rotation eigenstates from twelve-dimensional quantum calculations},
  author={Felker, Peter M and Ba{\v{c}}i{\'c}, Zlatko},
  journal={J. Chem. Phys.},
  volume={146},
  number={8},
  pages={084303},
  year={2017},
  publisher={AIP Publishing LLC},
  doi={10.1063/1.4976526}
}

@Article{belyanchikov2022single,
  author    = {Belyanchikov, MA and Bedran, ZV and Savinov, M and Bednyakov, P and Proschek, P and Prokleska, J and Abalmasov, VA and Zhukova, ES and Thomas, VG and Dudka, A and others},
  title     = {Single-particle and collective excitations of polar water molecules confined in nano-pores within a cordierite crystal lattice},
  journal   = {Phys. Chem. Chem. Phys.},
  year      = {2022},
  volume    = {24},
  number    = {11},
  pages     = {6890--6904},
  publisher = {Royal Society of Chemistry},
  doi       = {10.1039/d2cp03066g}
}

@Article{Serwatka2024,
  author        = {Serwatka, Tobias and Roy, Pierre-Nicholas},
  journal       = {J. Chem. Phys.},
  title         = {Quantum criticality in chains of planar rotors with dipolar interactions},
  year          = {2024},
  issn          = {1089-7690},
  month         = {mar},
  number        = {10},
  pages = {104302},
  volume        = {160},
  publisher     = {AIP Publishing},
  doi={https://doi.org/10.1063/5.0195453}
}

@Article{serwatka2022ferroelectric,
  author    = {Serwatka, Tobias and Roy, Pierre-Nicholas},
  title     = {Ferroelectric water chains in carbon nanotubes: Creation and manipulation of ordered quantum phases},
  journal   = {J. Chem. Phys.},
  year      = {2022},
  volume    = {157},
  number    = {23},
  pages     = {234301},
  publisher = {AIP Publishing LLC},
  doi       = {10.1063/5.0131149}
}

@Article{serwatka2022ground,
  author    = {Serwatka, Tobias and Roy, Pierre-Nicholas},
  title     = {Ground state of asymmetric tops with DMRG: Water in one dimension},
  journal   = {J. Chem. Phys.},
  year      = {2022},
  volume    = {156},
  number    = {4},
  pages     = {044116},
  publisher = {AIP Publishing LLC},
  doi       = {10.1063/5.0078770}
}

@Article{white1992density,
  author    = {White, Steven R},
  title     = {Density matrix formulation for quantum renormalization groups},
  journal   = {Phys. Rev. Lett.},
  year      = {1992},
  volume    = {69},
  number    = {19},
  pages     = {2863},
  publisher = {APS},
  doi       = {10.1103/physrevlett.69.2863}
}

@Article{schollwock2011density,
  author    = {Schollw{\"o}ck, Ulrich},
  title     = {The density-matrix renormalization group in the age of matrix product states},
  journal   = {Ann. Phys.},
  year      = {2011},
  volume    = {326},
  number    = {1},
  pages     = {96--192},
  publisher = {Elsevier},  doi       = {10.1016/j.aop.2010.09.012}
}

@article{fishman2022itensor,
  title={The ITensor software library for tensor network calculations},
  author={Fishman, Matthew and White, Steven and Stoudenmire, Edwin},
  journal={SciPost Phys.},
  volume={},
  pages={4},
  year={2022},
  doi={10.21468/scipostphyscodeb.4}
}

@mastersthesis{MoeedThesis,
    title    = {Path integral and qubit encoding techniques for quantum simulations of discrete planar rotor lattices},
    school   = {University of Waterloo},
    author   = {Moeed, Muhammad Shaeer},
    year     = {2025},
    type     = {{MSc} dissertation},
    doi={10012/21662},
    pages={28-29}
}

@mastersthesis{BoLiu,
    title    = {Mean-Field Theory of Collective Behavior of Quantum Dipolar Rotor},
    school   = {University of Waterloo},
    author   = {Liu, Bo},
    year     = {2021},
    type     = {{MSc} dissertation},
    doi={10012/16802},
    chapters={2-4}
}

@article{D_Frenkel_2001,
doi = {10.1088/0305-4470/34/17/302},
url = {https://doi.org/10.1088/0305-4470/34/17/302},
year = {2001},
month = {may},
publisher = {},
volume = {34},
number = {17},
pages = {3541},
author = {D Frenkel and R Portugal},
title = {Algebraic methods to compute Mathieu functions},
journal = {J. Phys. A: Math. Gen.}
}

@article{endohedral_fullerites_exp,
  title = {Endohedral fullerites: A new class of ferroelectric materials},
  author = {Cioslowski, Jerzy and Nanayakkara, Asiri},
  journal = {Phys. Rev. Lett.},
  volume = {69},
  issue = {19},
  pages = {2871--2873},
  numpages = {0},
  year = {1992},
  month = {Nov},
  publisher = {American Physical Society},
  doi = {10.1103/PhysRevLett.69.2871},
  OPTurl = {https://link.aps.org/doi/10.1103/PhysRevLett.69.2871}
}

@article{gorshunov_2016_incipient,
  author = {Gorshunov, B P and Torgashev, V I and Zhukova, E S and Thomas, Victor G and Belyanchikov, M A and Kadlec, Christelle and Kadlec, F and M. Savinov and T. Ostapchuk and J. Petzelt and Prokleška, Jan and Tomas, P V and Pestrjakov, Efim V and Fursenko, Dmitry A and Shakurov, G S and Prokhorov, A S and Gorelik, V S and Kadyrov, L S and Uskov, V V and Kremer, Reinhard K and Dressel, Martin},
  month = {09},
  publisher = {Nature Portfolio},
  title = {Incipient ferroelectricity of water molecules confined to nano-channels of beryl},
  volume = {7},
  pages={12822},
  year = {2016},
  journal = {Nat. Commun.},
  doi = {10.1038/ncomms12822}
}

@article{exp_nanotube,
author = {Biskupek, Johannes and Skowron, Stephen T. and Stoppiello, Craig T. and Rance, Graham A. and Alom, Shamim and Fung, Kayleigh L. Y. and Whitby, Richard J. and Levitt, Malcolm H. and Ramasse, Quentin M. and Kaiser, Ute and Besley, Elena and Khlobystov, Andrei N.},
title = {{B}ond {D}issociation and {R}eactivity of {HF} and {H$_2$O} in a {N}ano {T}est {T}ube},
journal = {ACS Nano},
volume = {14},
number = {9},
pages = {11178-11189},
year = {2020},
doi = {10.1021/acsnano.0c02661}
}

@article{hou2022synthesis,
  title={Synthesis of a monolayer fullerene network},
  author={Hou, Lingxiang and Cui, Xueping and Guan, Bo and Wang, Shaozhi and Li, Ruian and Liu, Yunqi and Zhu, Daoben and Zheng, Jian},
  journal={Nature},
  volume={606},
  number={7914},
  pages={507--510},
  year={2022},
  publisher={Nature Publishing Group UK London},
  doi={10.1038/s41586-022-04771-5}
}

@article{belyanchikov_2020_dielectric,
  author = {Belyanchikov, M. A. and Savinov, M. and Bedran, Z. V. and Bednyakov, P. and Proschek, P. and Prokleska, J. and Abalmasov, V. A. and Petzelt, J. and Zhukova, E. S. and Thomas, V. G. and Dudka, A. and Zhugayevych, A. and Prokhorov, A. S. and Anzin, V. B. and Kremer, R. K. and Fischer, H. and Lunkenheimer, P. and Loidl, A. and Uykur, E. and Dressel, M. and Gorshunov, B.},
  title = {Dielectric ordering of water molecules arranged in a dipolar lattice},
  volume = {11},
  pages={3927},
  year = {2020},
  journal = {Nat. Commun.},
  publisher = {Nature Portfolio},
  doi = {10.1038/s41467-020-17832-y}
}

@article{iouchtchenko2018ground,
  title={Ground states of linear rotor chains via the density matrix renormalization group},
  author={Iouchtchenko, Dmitri and Roy, Pierre-Nicholas},
  journal={J. Chem. Phys.},
  volume={148},
  number={13},
  pages={134115},
  year={2018},
  publisher={AIP Publishing LLC},
  doi={10.1063/1.5024403}
}

@Article{halverson2018quantifying,
  author    = {Halverson, Tom and Iouchtchenko, Dmitri and Roy, Pierre-Nicholas},
  title     = {Quantifying entanglement of rotor chains using basis truncation: Application to dipolar endofullerene peapods},
  journal   = {J. Chem. Phys.},
  year      = {2018},
  volume    = {148},
  number    = {7},
  pages     = {074112},
  publisher = {AIP Publishing LLC},
  doi       = {10.1063/1.5011769}
}

@article{Moed2024,
    author = {Moeed, Muhammad Shaeer and Serwatka, Tobias and Roy, Pierre-Nicholas},
    title = {Pair approximating the action for molecular rotations in path integral Monte Carlo},
    journal = {J. Chem. Phys.},
    volume = {162},
    number = {2},
    pages = {024113},
    year = {2025},
    month = {01},
    issn = {0021-9606},
    doi = {10.1063/5.0246327},
    url = {https://doi.org/10.1063/5.0246327}
}

@Article{Zhang2024,
    author = {Zhang, Wenxue and Moeed, Muhammad Shaeer and Bright, Andrew and Serwatka, Tobias and De Oliveira, Estevao and Roy, Pierre-Nicholas},
    title = {Path integral Monte Carlo in a discrete variable representation with Gibbs sampling: Dipolar planar rotor chain},
    journal = {J. Chem. Phys.},
    volume = {162},
    number = {1},
    pages = {014106},
    year = {2025},
    month = {01},
    issn = {0021-9606},
    doi = {10.1063/5.0244162},
    url = {https://doi.org/10.1063/5.0244162}
}

@Article{ma2017quasiphase,
  author    = {Ma, Xuedan and Cambr{\'e}, Sofie and Wenseleers, Wim and Doorn, Stephen K and Htoon, Han},
  title     = {Quasiphase transition in a single file of water molecules encapsulated in (6, 5) carbon nanotubes observed by temperature-dependent photoluminescence spectroscopy},
  journal   = {Phys. Rev. Lett.},
  year      = {2017},
  volume    = {118},
  number    = {2},
  pages     = {027402},
  publisher = {APS},  doi       = {10.1103/physrevlett.118.027402}}

@article{aoyagi2014cubic,
  title={A cubic dipole lattice of water molecules trapped inside carbon cages},
  author={Aoyagi, Shinobu and Hoshino, Norihisa and Akutagawa, Tomoyuki and Sado, Yuki and Kitaura, Ryo and Shinohara, Hisanori and Sugimoto, Kunihisa and Zhang, Rui and Murata, Yasujiro},
  journal={Chem. Comm.},
  volume={50},
  number={5},
  pages={524--526},
  year={2014},
  publisher={Royal Society of Chemistry},
  doi={https://doi.org/10.1039/C3CC46683C}
}

@article{Kastrup_2006,
  title = {Quantization of the canonically conjugate pair angle and orbital angular momentum},
  author = {Kastrup, H. A.},
  journal = {Phys. Rev. A},
  volume = {73},
  issue = {5},
  pages = {052104},
  numpages = {26},
  year = {2006},
  month = {May},
  publisher = {American Physical Society},
  doi = {10.1103/PhysRevA.73.052104},
  url = {https://link.aps.org/doi/10.1103/PhysRevA.73.052104}
}

@article{eh_ek_2008,
  year      = {2008},
  month     = {mar},
  publisher = {American Physical Society ({APS})},
  volume    = {77},
  number    = {3},
  pages={032110},
  author    = {J. {\v{R}}eh{\'{a}}{\v{c}}ek and Z. Bouchal and R. {\v{C}}elechovsk{\'{y}} and Z. Hradil and L. L. S{\'{a}}nchez-Soto},
  title     = {Experimental test of uncertainty relations for quantum mechanics on a circle},
  journal   = {Phys. Rev. A},
  doi       = {10.1103/physreva.77.032110}
}

@article{kleinert1990path,
  title={Path integral on spherical surfaces in D dimensions and on group spaces},
  author={Kleinert, H},
  journal={Phys. Lett. B},
  volume={236},
  number={3},
  pages={315--320},
  year={1990},
  publisher={Elsevier},
  doi={10.1016/0370-2693(90)90989-j}
}

@inbook{kleinert1993quantum,
  title = {Solution of Further Path Integrals by the {D}uru-{K}leinert Method},
  booktitle = {Path Integrals in Quantum Mechanics, Statistics, and Polymer Physics},
  author={Hagen Kleinert},
  month={june},
  year={1995},
  edition={2},
  chapter = {14},
  pages = {529--577},
  publisher={World Scientific Publishing},
  doi = {10.1142/9789814535519_0014}
}

@article{kleinert1997proper,
  title={Proper Dirac quantization of a free particle on a D-dimensional sphere},
  author={Kleinert, Hagen and Shabanov, Sergei V},
  journal={Phys. Lett. A},
  volume={232},
  number={5},
  pages={327--332},
  year={1997},
  publisher={Elsevier},
  doi={10.1016/S0375-9601(97)00410-6}
}

@unpublished{dolan2021constrained,
  title={Constrained Dynamics in the Hamiltonian formalism},
  author={Dolan, Brian P},
  note={arXiv:2112.12216v1 [hep-th]},
  year={2021},
  doi={10.48550/arXiv.2112.12216}
}

@article{Serwatka2023QPT,
  title = {Quantum Phase Transition in the One-Dimensional Water Chain},
  author = {Serwatka, T. and Melko, R. G. and Burkov, A. and Roy, P.-N.},
  journal = {Phys. Rev. Lett.},
  volume = {130},
  issue = {2},
  pages = {026201},
  numpages = {6},
  year = {2023},
  month = {Jan},
  publisher = {American Physical Society},
  doi = {10.1103/PhysRevLett.130.026201},
  url = {https://link.aps.org/doi/10.1103/PhysRevLett.130.026201}
}

@article{Serwatka2023endo,
author = {Serwatka, Tobias and Roy, Pierre-Nicholas},
title = {Quantum Criticality and Universal Behavior in Molecular Dipolar Lattices of Endofullerenes},
journal = {J Phys. Chem. Lett.},
volume = {14},
number = {24},
pages = {5586-5591},
year = {2023},
doi = {10.1021/acs.jpclett.3c01042}
}

@book{abramowitz1965handbook,
  title={Handbook of Mathematical Functions: With Formulas, Graphs, and Mathematical Tables},
  author={Abramowitz, M. and Stegun, I.A.},
  isbn={9780486612720},
  lccn={lc65012253},
  series={Applied mathematics series},
  url={https://books.google.ca/books?id=MtU8uP7XMvoC},
  year={1965},
  publisher={Dover Publications},
  chapter={17},
  doi={10.1115/1.3625776}
}

@article{PapoulisAthanasios1967PRVa,
author = {Papoulis, Athanasios and Hoffman, Joseph G.},
issn = {0031-9228},
journal = {Phys. today},
number = {1},
pages = {135-135},
title = {Probability, Random Variables, and Stochastic Processes},
volume = {20},
year = {1967},
doi={https://doi.org/10.1063/1.3034123}
}

@misc{winkelbauer2014momentsabsolutemomentsnormal,
      title={Moments and Absolute Moments of the Normal Distribution}, 
      author={Andreas Winkelbauer},
      year={2014},
      doi={10.48550/arXiv.1209.4340}
}

@Article{Dolgonos_2014,
author ="Dolgonos, Grygoriy A. and Peslherbe, Gilles H.",
title  ="Encapsulation of diatomic molecules in fullerene C60: implications for their main properties",
journal  ="Phys. Chem. Chem. Phys.",
year  ="2014",
volume  ="16",
issue  ="47",
pages  ="26294-26305",
publisher  ="The Royal Society of Chemistry",
doi  ="10.1039/C4CP04069D",
url  ="http://dx.doi.org/10.1039/C4CP04069D"}

@article{WOLFF199093,
title = {Critical slowing down},
journal = {Nucl. Phys. B - Proc. Supp.},
volume = {17},
pages = {93-102},
year = {1990},
issn = {0920-5632},
doi = {https://doi.org/10.1016/0920-5632(90)90224-I},
url = {https://www.sciencedirect.com/science/article/pii/092056329090224I},
author = {Ulli Wolff}
}

@article{Bonati2018,
  title = {Topological critical slowing down: Variations on a toy model},
  author = {Bonati, Claudio and D'Elia, Massimo},
  journal = {Phys. Rev. E},
  volume = {98},
  issue = {1},
  pages = {013308},
  numpages = {16},
  year = {2018},
  month = {Jul},
  publisher = {American Physical Society},
  doi = {10.1103/PhysRevE.98.013308},
  url = {https://link.aps.org/doi/10.1103/PhysRevE.98.013308}
}

@book{altland2010condensed,
  title={Condensed Matter Field Theory},
  author={Altland, A. and Simons, B.D.},
  isbn={9780521769754},
  lccn={2010282464},
  series={Cambridge books online},
  url={https://books.google.ca/books?id=GpF0Pgo8CqAC},
  year={2010},
  edition={2},
  publisher={Cambridge University Press},
  doi={10.1017/cbo9780511789984}
}

@article{wilkinson_approximate_2018,
title = {Approximate solutions to Mathieu's equation},
journal = {Physi. E: Low-Dimens. Syst. Nanostruct.},
volume = {100},
pages = {24-30},
year = {2018},
issn = {1386-9477},
doi = {https://doi.org/10.1016/j.physe.2018.02.019},
url = {https://www.sciencedirect.com/science/article/pii/S1386947717315990},
author = {Samuel A. Wilkinson and Nicolas Vogt and Dmitry S. Golubev and Jared H. Cole}
}

@Article{demille2002quantum,
  author    = {DeMille, David},
  title     = {Quantum computation with trapped polar molecules},
  journal   = {Phys. Rev. Lett.},
  year      = {2002},
  volume    = {88},
  number    = {6},
  pages     = {067901},
  publisher = {APS},
}

@article{Wei_2016,
author = {Wei, Qi and Cao, Yudong and Kais, Sabre and Friedrich, Bretislav and Herschbach, Dudley},
title = {Quantum Computation using Arrays of N Polar Molecules in Pendular States},
journal = {ChemPhysChem},
volume = {17},
number = {22},
pages = {3714-3722},
doi = {https://doi.org/10.1002/cphc.201600781},
year = {2016}
}

%%%% SUPPLEMENTARY MATERIAL %%%%
\onecolumngrid
\pagebreak
\clearpage

{
\sffamily
\bfseries
\linespread{1.5}
\Large 
\noindent
Supplementary Material for: Path Integral Monte Carlo in the Angular Momentum Basis for a Chain of Planar Rotors
}

%%%%%%%%%% Prefix a "S" to all equations, figures, tables and reset the counter %%%%%%%%%%
\setcounter{equation}{0}
\setcounter{figure}{0}
\setcounter{table}{0}
\setcounter{page}{1}
\setcounter{section}{0}
\makeatletter
\renewcommand{\theequation}{S\arabic{equation}}
\renewcommand{\thefigure}{S\arabic{figure}}
%%%%%%%%%% Prefix a "S" to all equations, figures, tables and reset the counter %%%%%%%%%%

\section{Taylor expansion of the dipole-dipole potential}
\label{ap:Taylor_Expansion_Potential}
The equilibrium points $\va{\theta} = \left(\theta_1,\theta_2,\dots, \theta_N\right)$ are given by the solution of the equation
\begin{align}
\left[\grad{V(\va{\varphi})}\right]_{\va{\varphi}= \va{\theta}} 
=g\sum_{i=1}^{N} \vb{e}_i \left[\left( \sin{\theta_{i-1}} + \sin{\theta_{i+1}}\right)\cos{\theta_i} + 2 \left( \cos{\theta_{i-1}} + \cos{\theta_{i+1}}\right)\sin{\theta_i}\right]
=0,
\end{align}
implying that $\theta_i = \theta \equiv \frac{n\pi}{2}$, for all $i=1,2,\dots$, and $n=0,1,2,\dots$. By inspection of the second derivatives of $V(\va{\varphi})$ we can see that $n$ even (odd) gives the stable (unstable) equilibrium points. Therefore, a Taylor expansion of $V(\va{\varphi})$ around its equilibrium points is
\begin{align}
V(\va{\varphi}) 
&\approx V(\theta) + \frac{1}{2} \sum_{j,k} \left[\mathbf{H}(\theta)\right]_{ij} \xi_j \xi_k
\nn
&= -g N f_{\theta} + g\sum_{j} f_{\theta}\xi_{j}^2 + \frac{f_{\theta}-1}{2}\left(\xi_{j-1} + \xi_{j+1}\right)\xi_{j},
\end{align}
where $\mathbf{H}$ is the Hessian matrix of the potential, $\xi_{j} \equiv \varphi_j - \theta$ are the small oscillations around the equilibrium point, and $f_{\theta} \equiv  3\cos^2{\theta} - 1$.

\section{Derivation of the kinetic and potential energies in the reciprocal lattice coordinates}
\label{ap:normal_modes}

For the Kinetic energy, substituting Eq. (7) into Eq. (5) leads to
\begin{align}
	\label{eq:kinetic_SOA_NM}
	K 
	= \frac{I}{2} \sum_i \dot{\xi_i}^2
	= \frac{I}{2} \sum_i \frac{1}{N} \sum_{j,k} \dot{\phi}_j \dot{\phi}_k \exp{\i \frac{2\pi}{N}i(j+k)}
	= \frac{I}{2} \sum_{j} \dot{\phi}_j {\dot{\phi}}_{-j}.
	= \frac{I}{2} \sum_{j} \dot{\phi}_j \cc{\dot{\phi}}_j.
\end{align}
For the potential energy, substituting Eq. (7) into Eq. (4) leads to
\begin{align}
	\label{eq:potential_SOA_NM}
	V
	&=-2gN + g \sum_{i} \left(2 \xi_{i}^2  + \frac{\xi_{i-1}\xi_i + \xi_{i+1}\xi_i}{2}\right)
	\nn
	&=-2gN +g \sum_{i} \left(
		\frac{2}{N} \sum_{j,k} \phi_j \phi_k \exp{\i \frac{2\pi}{N}i(j+k)}
		+ \frac{1}{2N}\sum_{j,k} \phi_j \phi_k \exp{\i \frac{2\pi}{N}(i-1)j}\exp{\i \frac{2\pi}{N}ik}
		+ \frac{1}{2N}\sum_{j,k} \phi_j \phi_k \exp{\i \frac{2\pi}{N}(i+1)j}\exp{\i \frac{2\pi}{N}ik}
	\right)
	\nn 
	&=-2gN +g \sum_{j,k}\frac{1}{N} \sum_{i} \exp{\i \frac{2\pi}{N}i(j+k)}\left(2 + \frac{\exp{\i \frac{2\pi}{N}j}+\exp{-\i \frac{2\pi}{N}j}}{2}\right) \phi_j \phi_k
	\nn 
	&=-2gN +g \sum_{j} \left[2 + \cos({\textstyle \frac{2\pi}{N}}j)\right] \phi_j \cc{\phi}_{j}
	.
\end{align}
It is important to notice that for the case where $N=2$ the substitution $g \rightarrow \frac{g}{2}$ needs to be made to avoid the double counting in the sums.

\section{Spectral decomposition and canonical variable transformation of the Classical Hamiltonian in the NM coordinates}
\label{ap:spectral_decomp}

From Eq. (9)
\begin{align}
	\phi_j &= a_j \exp{\i \omega_j t}+ \cc{a}_{-j} \exp{-\i \omega_j t} ,
	\\
	\dot{\phi}_j &= \i \omega_j \left( a_j \exp{\i \omega_j t} - \cc{a}_{-j} \exp{-\i \omega_j t} \right) ,
\end{align}
we have that
\begin{equation}
	\omega_j^2\phi_j\cc{\phi}_j 
	=\omega_j^2 \left(a_j \exp{\i \omega_j t}+ \cc{a}_{-j} \exp{-\i \omega_j t}\right)\left(\cc{a}_j \exp{-\i \omega_j t}+ {a}_{-j} \exp{\i \omega_j t}\right)
	= \omega_j^2 \left(a_j \cc{a}_{j} + a_{-j} \cc{a}_{-j} + {a}_{j}{a}_{-j} \exp{2\i \omega_j t} + \cc{a}_{-j}\cc{a}_{j} \exp{-2\i \omega_j t}\right),
\end{equation}
and
\begin{equation}
	\dot{\phi}_j\cc{\dot{\phi}}_j 
	= \omega_j^2 \left(a_j \exp{\i \omega_j t} - \cc{a}_{-j} \exp{-\i \omega_j t}\right)\left(\cc{a}_j \exp{-\i \omega_j t} - {a}_{-j} \exp{\i \omega_j t}\right)
	= \omega_j^2 \left(a_j \cc{a}_{j} + a_{-j} \cc{a}_{-j} - {a}_{j}{a}_{-j} \exp{2\i \omega_j t} - \cc{a}_{-j}\cc{a}_{j} \exp{-2\i \omega_j t}\right),
\end{equation}
implying that
\begin{align}
	\frac{I}{2}\sum_j \dot{\phi}_j\cc{\dot{\phi}}_j + \omega_j^2\phi_j\cc{\phi}_j
	= \sum_j \frac{I\omega_j^2}{2} \left(2a_j \cc{a}_{j} + 2a_{-j} \cc{a}_{-j} \right)
	= \sum_j 2I \omega_j^2\, a_j \cc{a}_{j}
	.
\end{align}
Now, by defining $\{Q_j, P_j\}$ as the generalized coordinates of position and momentum
\begin{align}
	Q_j = \gamma \left(a_j \exp{\i \omega_j t} + \cc{a}_{j} \exp{-\i \omega_j t} \right) ,
	\\ 
	P_j = \dot{Q}_j = \gamma \i \omega_j \left(a_j \exp{\i \omega_j t} - \cc{a}_{j} \exp{-\i \omega_j t} \right).
\end{align}
Then,
\begin{align}
	\frac{P_j^2}{2} + \frac{\omega^2_j}{2} Q^2_j
    &= - \frac{\gamma^2  \omega^2_j}{2} \left(a_j \exp{\i \omega_j t} - \cc{a}_{j} \exp{-\i \omega_j t}  \right)^2 + \frac{\gamma^2 \omega^2_j}{2}\left(a_j \exp{\i \omega_j t} + \cc{a}_{j} \exp{-\i \omega_j t}  \right)^2 
    \nn
	&= - \frac{\gamma^2  \omega^2_j}{2} \left(- 2 a_j\cc{a}_{j} +  a^2_j \exp{2\i \omega_j t} + (\cc{a}_{j})^2 \exp{-2\i \omega_j t} \right) + \frac{\gamma^2 \omega^2_j}{2} \left(2 a_j\cc{a}_{j} +  a^2_j \exp{2\i \omega_j t} + (\cc{a}_{j})^2 \exp{-2\i \omega_j t} \right)
    \nn
	&= 2 \gamma^2 \omega_j^2 a_j\cc{a}_{j}.
\end{align}
Therefore, in order to write the spectral decomposition of the harmonic part in the form of Eq. (10), keeping the consistency of the canonical equations of motion, we have that $2\gamma^2 = 2I \implies \gamma =  \sqrt{I}$. The position and momentum generalized coordinates then become
\begin{align}
	Q_j = \sqrt{I} \left(a_j \exp{\i \omega_j t} + \cc{a}_{j} \exp{-\i \omega_j t} \right) \, ,
	\\ 
	P_j = \i \omega_j \sqrt{I} \left(a_j \exp{\i \omega_j t} - \cc{a}_{j} \exp{-\i \omega_j t} \right) .
\end{align}
Therefore, we can now write the position and momentum generalized coordinates in terms of the original coordinates of the planar rotors system
\begin{align}
	Q_j
	&= \sqrt{I} \left(\frac{\phi_j + \cc{\phi}_j}{2} + \frac{\dot{\phi}_j-\cc{\dot{\phi}}_j}{2\i \omega_j}\right),
	\nn
	P_j
	&= \sqrt{I} \i \omega_j \left(\frac{\phi_j - \cc{\phi}_j}{2} + \frac{\dot{\phi}_j+\cc{\dot{\phi}}_j}{2\i \omega_j}\right)
	,
\end{align}
implying that 
\begin{align}
	\label{eq:relationP,Q_phidot,phi}
	\frac{P_j^2}{2} + \frac{\omega^2_j}{2} Q^2_j
	=
	\frac{I}{2}\left|\dot{\phi}_j\right|^2 + \frac{I\omega^2_j}{2} \left|\phi_j\right|^2
	.
\end{align}

\section{Second quantization process}
\label{ap:second_quatization}
Introducing the ladder operators for the QHO
\begin{align}
	\oper{a}_j &= \sqrt{ \frac{\omega_j}{2\hbar }}\left(\oper{Q}_j + \frac{\i}{\omega_j} \oper{P}_j\right)
	,
	&
	\hc{\oper{a}}_j &=\sqrt{ \frac{\omega_j}{2 \hbar}} \left(\oper{Q}_j - \frac{\i}{\omega_j} \oper{P}_j\right)
\end{align}
which implies
\begin{align}
	\oper{Q}_j &= \sqrt{ \frac{\hbar}{2\omega_j}}  \left(\hc{\oper{a}}_j + \oper{a}_j\right)
	,
	&
	\oper{P}_j &= \i \sqrt{ \frac{\hbar \omega_j}{2}} \left(\hc{\oper{a}}_j - \oper{a}_j\right),
\end{align}
keeping the canonical commutation relations consistent
\begin{align}
    \comm{\oper{Q}_j}{\oper{P}_j} = \i \iff \comm{\oper{a}}{\hc{\oper{a}}} = 1
    .
\end{align}
Then,
\begin{align}
    \frac{\oper{P}_j^2}{2} + \frac{\omega_j^2}{2}\oper{Q}_j^2
    &=
    -\frac{\hbar \omega_j}{4} \left(\hc{\oper{a}}_j - \oper{a}_j\right)^2 + \frac{\omega_j}{4} \left(\hc{\oper{a}}_j + \oper{a}_j\right)^2
    \nn
    &=
    -\frac{\hbar \omega_j}{4} \left[(\hc{\oper{a}}_j)^2 + ({\oper{a}}_j)^2  - \oper{a}_j \hc{\oper{a}}_j - \hc{\oper{a}}_j\oper{a}_j  \right] + \frac{\omega_j}{4} \left[(\hc{\oper{a}}_j)^2 + ({\oper{a}}_j)^2  + \oper{a}_j \hc{\oper{a}}_j + \hc{\oper{a}}_j\oper{a}_j \right]
    \nn
    &=
    \frac{\hbar \omega_j}{2} \left(2\hc{\oper{a}}_j \oper{a}_j + 1\right)
    \nn
    &=\hbar \omega_j \left(\hc{\oper{a}}_j \oper{a}_j + \frac{1}{2}\right).
\end{align}

\section{Physical observables of interest in the ordered phase}
\label{ap:expval_ordered_phase}

\subsection{Polarization in the x-axis direction}S

Here the total polarization operator can be defined as
\begin{align}
	\obs{M} 
	= \sum_i \cos{\varphi}_i
	&\approx  \sum_i \cos{\theta_i} \left(1-\frac{\xi_i^2}{2}\right) 
	\nn
	&= N\cos{\theta}  -\frac{\cos{\theta}}{2} \frac{1}{N} \sum_i \sum_j \sum_k \phi_j \phi_k e^{\i\frac{2\pi}{N} (j+k)i}
	\nn
	&= N\cos{\theta}  -\frac{\cos{\theta}}{2} \sum_j \phi_j \cc{\phi}_j 
	.
\end{align}
Now, for the quantum case we can make the substitution $\phi_j \cc{\phi}_j \rightarrow \oper{\phi}_j \hc{\oper{\phi}}_j \rightarrow \frac{\oper{Q}_j^2}{I}$, resulting in
\begin{align}
	\obsop{M}
	= \cos{\theta}  \left(N - \sum_j \frac{\oper{Q}_j^2}{2I}\right) 
	.
\end{align}

\subsection{Orientational correlation}

Here the orientational correlation operator can be defined as
\begin{align}
	\obs{C}
	&= \sum_i \cos(\varphi_i - \varphi_{i+1})
	\nn
	&= \sum_i \cos{\varphi}_i\cos{\varphi}_{i+1} + \sin{\varphi}_i\sin{\varphi}_{i+1}
	\nn
	&\approx \cos^2{\theta} \sum_i \left(1-\frac{\xi_i^2}{2}\right)\left(1-\frac{\xi_{i+1}^2}{2}\right) + \xi_{i}\xi_{i+1}
	\nn
	&= N - \frac{1}{2} \sum_{i} \left(\xi_i - \xi_{i+1}\right)^2
	\nn
	&= N - \frac{1}{2N} \sum_{i} \sum_{j,k} \phi_j\phi_k e^{\i \frac{2\pi}{N}(j+k)i} \left(1 - e^{\i \frac{2\pi}{N}j}\right)\left(1 - e^{\i \frac{2\pi}{N}k}\right)
	\nn
	&= N - \frac{1}{2} \sum_{j} \phi_j\cc{\phi}_j \left(1 - e^{\i \frac{2\pi}{N}j}\right)\left(1 - e^{-\i \frac{2\pi}{N}j}\right)
	\nn
	&= N + \sum_{j} \left(-1 + \cos({\textstyle \frac{2\pi}{N}}j)\right)\phi_j\cc{\phi}_j 
	.
\end{align}
Again, for the quantum case we can make the substitution $\phi_j \cc{\phi}_j \rightarrow \oper{\phi}_j \hc{\oper{\phi}}_j \rightarrow \frac{\oper{Q}_j^2}{I}$, resulting in
\begin{align}
	\obsop{C}
	= N + \sum_{j} \left(-1 + \cos({\textstyle \frac{2\pi}{N}}j)\right)\frac{\oper{Q}_j^2}{I} 
	.
\end{align}

\section{Asymptotic analysis of the quartic interactions}
\label{ap:asymp_expan}

For the Hamiltonian of form
\begin{align}
    H = H_{\text{HO}} + g\sum_{\alpha>2} \{\phi^{\alpha}\},
\end{align} 
where $\{\phi^{\alpha}\}$ stands for the contribution of all $\alpha$-th order terms in the expansion, for $\alpha$ even, and $H_{\text{HO}}$ is the QHO Hamiltonian for all NM.
\begin{align}
	\oper{H}_{\text{HO}} = \sum_j \oper{H}_{j}^{(0)},
\end{align}
for
\begin{align}
	\oper{H}_{j}^{(0)} \ket{n_j}.
\end{align}
Then, the first order correction to the energy due to $\{\phi^{\alpha}\}$ is
\begin{equation}
	\varepsilon^{(1)}_j 
	\propto g \mel{n_j}{\{\phi^{\alpha}\}_j}{ n_j}
	\propto g \mel{n_j}{Q_j^{\alpha}}{ n_j}
	,
\end{equation}
where the definitions from Sec. \ref{ap:spectral_decomp} were used in the last line. For the ground state, i.e. $n_j=0 \forall j$, $\mel{0_j}{\{\phi^{\alpha}\}_j}{0_j} \equiv \expval{Q_j^{\alpha}}$ is simply the $\alpha$-th moment of the Gaussian distribution defined by the ground state vector of the $j$-th mode. Defining $\equiv \expval{Q_j^{\alpha}}$ in terms of the mean $\expval{Q_j}=0$ and variance $\expval{Q_j^2}=\frac{1}{4\kappa_j\sqrt{g}}$ of the distribution gives
\begin{align}
	\expval{Q_j^{\alpha}}
	= \expval{Q_j^2}^\frac{\alpha}{2} (\alpha -1)!!
	,
\end{align}
where $!!$ represents the double factorial.\cite{PapoulisAthanasios1967PRVa,winkelbauer2014momentsabsolutemomentsnormal} Therefore, the first order correction in the ground state energy for the $j$-th mode will be
\begin{equation}
	\varepsilon^{(1)}_j 
	\propto g \expval{Q_j^{\alpha}}
	\propto g \expval{Q_j^2}^\frac{\alpha}{2}
	\propto g \, g^{-\frac{\alpha}{4}} = g^{\frac{4-\alpha}{4}}
	.
\end{equation}
Analogously, the first order corrections in the ground state vector is
\begin{align}
	\ket*{\psi_j^{(1)}}
	= g \sum_{n_j\neq 0} \frac{\mel*{n_j}{\{\phi^{\alpha}\}_j}{0_j}}{E_{0}^{(0)} - E_{n_j}^{(0)}}\ket{n_j}
	\propto 
	\frac{g \, g^{-\frac{\alpha}{4}}}{g^{\frac{1}{2}}} \sum_{n_j \neq 0} \lambda_{n_j} \ket{n_j}
	=
	g^{\frac{2-\alpha}{4}} \sum_{n_j \neq 0} \lambda_{n_j} \ket{n_j},
\end{align}
for $\lambda_{n_j}$ probability amplitude.

\section{Perturbation Theory for the disordered phase}
\label{ap:pert_theo_disorder}

The correction in the ground state energy, up to second order, can be calculated as follows
\begin{align}
	E_0 
	&= E_0^{(0)} + g \expval*{\V{}}{\va{0}} + g^2 \sum_{\va{m}\neq \va{0}} \frac{\big|\mel*{\va{m}}{\V{}}{\va{0}}\big|^2}{E_{\va{0}}^{(0)} - E_{\va{m}}^{(0)}}
	\nn
	&= - g^2 \sum_{\va{m}\neq 0}  \frac{\sum_i \abs{\mel*{m_i,m_{i+1}}{\V{i,i+1}}{0,0}}^2}{\sum_i m_i^2}
	\nn
	&= - \frac{g^2}{16} \sum_{\va{m}\neq 0}  \frac{\sum_i \left(
		3\delta_{m_i,1}\delta_{m_{i+1},1}
		+\delta_{m_i,-1}\delta_{m_{i+1},1}
		+\delta_{m_i,1}\delta_{m_{i+1},-1}
		+3\delta_{m_i,-1}\delta_{m_{i+1},-1}
		\right)^2 \prod_{j \neq i,i+1} \delta_{m_j,0}}{\sum_i m_i^2}
	\nn
	&= - \frac{g^2}{16} \sum_i  \frac{3^2}{1^2 + 1^2 }  + \frac{1}{(-1)^2 + 1^2 } + \frac{1}{(1)^2 + (-1)^2 } + \frac{3^2}{(-1)^2 + (-1)^2 }
	\nn
	&= - \frac{20}{32} g^2 = - \frac{5}{8} g^2 N
	,
\end{align}
where ${\sum}^{\ast}$ refers to the condition $\va{m} \neq 0$. 
In addition, the correction in the ground state vector, up to first order, can be calculated using the angular momentum basis representation as follows
\begin{align}
	\ket{\psi_0}
	&= \ket*{\va{0}}  + g \sum_{\va{m}\neq \va{0}} \frac{\mel*{\va{m}}{\V{}}{\va{0}}}{E_{\va{0}}^{(0)} - E_{\va{m}}^{(0)}}\ket{\va{m}}
	\nn
	&=\ket*{\va{0}}   - g \sum_{\va{m}\neq 0}  \frac{\sum_i \mel*{m_i,m_{i+1}}{\V{i,i+1}}{0,0}}{\sum_i m_i^2} \ket{\va{m}}
	\nn
	&=\ket*{\va{0}}  + \frac{g}{4} \sum_{\va{m}\neq 0}  \frac{\sum_i \left(
		3\delta_{m_i,1}\delta_{m_{i+1},1}
		+\delta_{m_i,-1}\delta_{m_{i+1},1}
		+\delta_{m_i,1}\delta_{m_{i+1},-1}
		+3\delta_{m_i,-1}\delta_{m_{i+1},-1}
		\right) \prod_{j \neq i,i+1} \delta_{m_j,0}}{\sum_i m_i^2} \ket{\va{m}}
	\nn
	&= \ket*{\va{0}}  + \frac{g}{4} \sum_i 
		\left[\frac{3\ket{1_i,1_{i+1}}}{1^2 + 1^2 }  + \frac{\ket{-1_i,1_{i+1}}}{(-1)^2 + 1^2 } + \frac{\ket{1_i,-1_{i+1}}}{(1)^2 + (-1)^2 } + \frac{3\ket{-1_i,-1_{i+1}}}{(-1)^2 + (-1)^2 }\right]\bigotimes_{j \neq i,i+1}\ket{0_j}
	\nn
	&= \ket*{\va{0}}  + \frac{g}{8} \sum_i 
		\left(3\ket{1_i,1_{i+1}}  + \ket{-1_i,1_{i+1}}+ \ket{1_i,-1_{i+1}}+ 3\ket{-1_i,-1_{i+1}}\right)
	,
\end{align}
where the tensor product $\bigotimes_{j \neq i,i+1}\ket{0_j}$ was omitted in the last line for the sake of notation. 
Now we can calculate the expectation value of the following physical properties:
\subsection{Variance of the total Angular Momentum}
\begin{align}
	\expval{\obsop{L}^2} 
	&= \expval{\sum_j \L{j} \sum_k \L{k}}{\psi_0}
	\nn
	&= \bra{\psi_0}\sum_j \L{j} \left[\frac{g}{8} \sum_i 
		\left(6\ket{1_i,1_{i+1}}  - 6 \ket{-1_i,-1_{i+1}}\right)\right]
	\nn
	&= \bra{\psi_0}\left[\frac{12g}{8} \sum_i 
		\left(\ket{1_i,1_{i+1}}  + \ket{-1_i,-1_{i+1}}\right)\right]
	\nn
	&= \left[\bra*{\va{0}}  + \frac{g}{8} \sum_i 
		\left(3\bra{1_i,1_{i+1}}  + \bra{-1_i,1_{i+1}}+ \bra{1_i,-1_{i+1}}+ 3\bra{-1_i,-1_{i+1}}\right)\right]\left[\frac{12g}{8} \sum_i 
		\left(\ket{1_i,1_{i+1}}  + \ket{-1_i,-1_{i+1}}\right)\right]
	\nn
	&= \frac{12^2}{64} \sum_i 
		\left(3\braket{1_i,1_{i+1}}  + 3\braket{-1_i,-1_{i+1}}\right)
	\nn
	&= \frac{72g^2}{64} N = \frac{9g^2}{8} N
	.
\end{align}

\subsection{Polarization in the x-axis direction}

Here the total polarization operator can be defined as
\begin{align}
	\obsop{M}
	= \sum_i \cos{\oper{\varphi}_i}
	= \frac{1}{2}\sum_i \E{+}{i} + \E{-}{i} ,
\end{align}
where in the last expression it was represented in the angular momentum basis.\cite{deOliveira2025} Then,
\begin{align}
	\expval{\obsop{M}} 
	&= \expval{\left(\frac{1}{2}\sum_i \E{+}{i} + \E{-}{i}\right)}{\psi_0}
	\nn
	&= \left(\bra{\psi_0^{(0)}} + g\bra{\psi_0^{(1)}}\right)\left(\sum_i \E{+}{i} + \E{-}{i}\right)  \left(\ket{\psi_0^{(0)}} + g\ket{\psi_0^{(1)}}\right),
\end{align}
and since $\ket*{\psi_0^{(0)}}$ contains zero angular momentum excitations, and $\ket*{\psi_0^{(1)}}$ contains a pair of angular momentum excitations, we have that
\begin{align}
	\mel**{\psi_0^{(0)}}{\E{\pm}{i}}{\psi_0^{(0)}}
	= \mel**{\psi_0^{(1)}}{\E{\pm}{i}}{\psi_0^{(0)}}
	= \mel**{\psi_0^{(1)}}{\E{\pm}{i}}{\psi_0^{(1)}}
	= 0
\end{align}

\subsection{Orientational correlation}

Here the orientational correlation operator can be defined as
\begin{align}
	\obsop{C}
	&= \sum_i \cos(\oper{\varphi}_i - \oper{\varphi}_{i+1})
	\nn
	&= \sum_i \cos{\oper{\varphi}_i}\cos{\oper{\varphi}_{i+1}} + \sin{\oper{\varphi}_i}\sin{\oper{\varphi}_{i+1}}
	\nn
	&= \sum_i \frac{\E{+}{i} + \E{-}{i}}{2}\frac{\E{+}{i+1} + \E{-}{i+1}}{2} + \frac{\E{+}{i} - \E{-}{i}}{2\i}\frac{\E{+}{i+1} - \E{-}{i+1}}{2\i}
	\nn
	&= \frac{1}{2}\sum_i \E{+}{i}\E{-}{i+1} + \E{-}{i}\E{+}{i+1} ,
\end{align}
where in the last expressions it was represented in the angular momentum basis.\cite{deOliveira2025}
Then,
\begin{align}
	\expval{\obs{C}} 
	&= \expval{\left(\frac{1}{2}\sum_i \E{+}{i}\E{-}{i+1} + \E{-}{i}\E{+}{i+1}\right)}{\psi_0}
	\nn
	&= \frac{1}{2}\sum_i \bra{\psi_0} \left( \E{+}{i}\E{-}{i+1} + \E{-}{i}\E{+}{i+1}\right)  \left[\ket*{\va{0}}  + \frac{g}{8} \sum_i 
		\left(3\ket{1_i,1_{i+1}}  + \ket{-1_i,1_{i+1}}+ \ket{1_i,-1_{i+1}}+ 3\ket{-1_i,-1_{i+1}}\right)\right],
	\nn
	&= \frac{g}{16}\sum_i \left( \mel{0_i,0_{i+1}}{\E{+}{i}\E{-}{i+1}}{-1_i,1_{i+1}} +\mel{1_i,-1_{i+1}}{\E{+}{i}\E{-}{i+1}}{0_i,0_{i+1}} + \text{h.c.}\right)
	\nn
	&= \frac{g}{4}N
	.
\end{align}
It is important to notice that for the case where $N=2$ the substitution $g \rightarrow \frac{g}{2}$ needs to be made to account for the double counting in the summations $\sum_i$ for all the calculated estimators above.

\end{document}